\documentclass[twocolumn,lengthcheck,superscriptaddress]{revtex4-1}

\usepackage{lmodern}
\usepackage{amsmath}
\usepackage{amsthm}
\usepackage{amssymb}
\usepackage{amsfonts}
\usepackage{graphicx}
\usepackage{hyperref}
\usepackage{color}
\usepackage{braket}

\newcommand{\tr}{\operatorname{tr}}

\newcommand{\sgn}{\operatorname{sgn}}
\newcommand{\cwplaqsub}{\vcenter{\hbox{\includegraphics{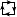}}}}

\def\su2{\textsl{SU}(2)}
\def\o3{\textsl{O}(3)}
\def\uone{\textsl{U}(1)}
\def\u2{\textsl{U}(2)}
\def\CL{\textsl{CL}}
\def\CR{\textsl{CR}}
\def\CU{\textsl{CU}}

\def\gl2c{\textsl{GL}(2, \mathbb{C})}
\def\lasu2{\mathfrak{su}(2)}

\theoremstyle{definition}

\theoremstyle{remark}

\begin{document}

\title{Quantum Yang-Mills theory: an overview of a programme}

\author{Ashley Milsted}
\affiliation{Perimeter Institute for Theoretical Physics, Waterloo, Ontario N2L 2Y5, Canada}
\affiliation{Institut f\"ur Theoretische Physik, Leibniz Universit\"at Hannover, Appelstr. 2, 30167 Hannover, Germany}

\author{Tobias J.\ Osborne}
\affiliation{Institut f\"ur Theoretische Physik, Leibniz Universit\"at Hannover, Appelstr. 2, 30167 Hannover, Germany}

\begin{abstract}
We present an overview of a programme to understand the low-energy physics of quantum Yang-Mills theory from a quantum-information perspective. Our setting is that of the hamiltonian formulation of pure Yang-Mills theory in the temporal gauge on the lattice. Firstly, inspired by recent constructions for $\mathbb{Z}/2\mathbb{Z}$ lattice gauge theory, in particular, Kitaev's toric code, we describe the gauge-invariant sector of hilbert space by introducing a primitive quantum gate:  the quantum parallel transporter. We then develop a nonabelian generalisation of laplace interpolation to present an ansatz for the ground state of pure Yang-Mills theory which interpolates between the weak- and strong-coupling RG fixed points. The resulting state acquires the structure of a tensor network, namely, a multiscale entanglement renormalisation ansatz, and allows for the efficient computation of local observables and Wilson loops. Various refinements of the tensor network are discussed leading to several generalisations. Finally, the continuum limit of our ansatz as the lattice regulator is removed is then described. This paper is intended as an abstract for an ongoing programme: there are still many open problems.
\end{abstract}

\maketitle

\section{Introduction}
Nonabelian gauge theory, known as Yang-Mills theory, is a fundamental component of the standard model of particle physics describing the dynamics of all known subatomic particles. Thanks to asymptotic freedom we now have a rather satisfactory understanding of the high-energy limit of Yang-Mills theory via perturbation theory. However, the non-perturbative infrared limit relevant for most observable physics has resisted complete solution. 

A most successful tool so far in the study of Yang-Mills theory and the standard model has been the computer. When quantum field theory is regulated (after a Wick rotation) on a space-time lattice \cite{wilson:1974b, creutz:1985a} the intractable path integral representation becomes amenable to Monte Carlo sampling. This approach has lead to unparalleled insights culminating in the recent determination of the hadronic spectrum of QCD \cite{duerr:2008a}. 

However, the success of Monte Carlo methods in the study of Yang-Mills theory and the standard model is not entirely satisfactory. Vast computational effort is required because the lattices  involve hundreds of thousands of sites and, consequently, many samples are necessary to reduce statistical errors. The brute-force approach of Monte Carlo is also somewhat at odds with our aesthetic desire for a ``physical'' understanding. Many believe that the  symmetry of Yang-Mills theory should result in a succinct explanation of its low-energy physics. These aspirations are best summarised by a quote of Polyakov\footnote{\url{http://quantumfrontiers.com/2012/12/11/fundamental-physics-prize-prediction-polyakov/}.}: \begin{quote} QCD must be exactly soluble, or else I cannot imagine what the physics textbooks of the future will look like. \end{quote} The search for a simpler explanation of the low-energy physics of Yang-Mills theory is a core motivation to consider approaches other than Monte Carlo.

Impressive analytic progress towards understanding the low-energy physics of pure Yang-Mills theory has been made in the past half century \cite{thooft:2005}. However, there are still some mysteries that deserve further investigation. One important problem is to describe, nonperturbatively, the continuum limit of the ground-state in the \emph{strong-coupling limit} and then to efficiently extract its large-scale behaviour. To do this we would like a compellingly simple ground-state ansatz which efficiently captures the large-scale physics of pure Yang-Mills theory while providing a parsimonious explanation for important features such as the area law behaviour of Wilson loops. 

Several notable ansatz ground states for pure Yang-Mills theory have been proposed in the past decades \cite{greensite:1979a, feynman:1981a, karabali:1998a, samuel:19967a}. These proposals variously take a \emph{dimensionally reduced} form and interpolate between the zero-coupling limit, where the ground state is a collection of copies of  electromagnetism, and the strong-coupling limit. Unfortunately all the proposals so far are not efficiently contractible and one must take recourse to Monte Carlo sampling. Hence we should still explore other descriptions of the ground state.

It is the purpose of the paper to propose such an ansatz: we argue here that the ground state of pure Yang-Mills theory finds its most economical representation as a \emph{tensor network state} (TNS). Further, this TNS is efficiently contractible and affords the efficient calculation of $n$-point correlation functions and Wilson loops without recourse to Monte Carlo sampling or perturbation theory.

Our study takes its inspiration from quantum-information theoretic driven progress in condensed matter physics. Here the variational method, combined with expressive variational classes, has proved to be a powerful tool in our understanding of the strongly correlated physics of \emph{quantum spin systems} \cite{auerbach:1994a, sachdev:2011a}. These variational approaches fall under the rubric of the \emph{density matrix renormalisation group} (DMRG) \cite{schollwoeck:2005a,schollwock:2011a} and have led to remarkable insights in recent years providing new tools to overcome many previously insurmountable roadblocks such as the simulation of dynamics \cite{vidal:2003a, haegeman:2011b} and fermions \cite{corboz:2009a, corboz:2010a, corboz:2010b, kraus:2010a} without sign problems and the determination of spectral information \cite{haegeman:2012a}. These developments are due, in no small part, to new impetus from quantum information theory in the understanding of \emph{quantum entanglement}. New entanglement-inspired tensor network state variational classes, including the projected entangled-pair states (PEPS) \cite{verstraete:2004a} and the multiscale entanglement renormalisation ansatz (MERA) \cite{vidal:2006a, vidal:2007a} have led to major progress in our understanding of strongly correlated phenomena.

There are now several crucial hints that tensor networks might be a powerful tool in the study of lattice gauge theory (for a review, see \cite{dalmonte_lattice_2016}). Firstly, several recent studies have applied one-dimensional TNS, i.e., matrix product states (MPS), to the Schwinger model with very encouraging results, including the determination of the particle content and real time evolution \cite{byrnes:2002a,sugihara:2005a,banuls:2013a,buyens:2014,rico:2014,pichler_real-time_2016, banuls_density_2017, buyens_finite-representation_2017-1}, as well as to non-abelian models \cite{kuhn:2015, silvi_finite-density_2017, banuls_efficient_2017}. Related work is also ongoing in the field of quantum gravity \cite{dittrich_decorated_2016, delcamp_phase_2017}. Secondly, the ground state space of $\mathbb{Z}/2\mathbb{Z}$-lattice gauge theory (and associated quantum double models) admits an efficient exact description as a TNS, namely a PEPS or a MERA, \cite{dennis:2002a, aguado:2008a}, a result later generalised to include string-net models \cite{buerschaper:2009a, koenig:2009a}. This construction has been supplemented with numerical results \cite{tagliacozzo:2011a} strongly indicating the utility of the MERA ansatz in the description of the low-energy physics of lattice gauge theories. These results are rather suggestive that an economic description of the low-energy limit of Yang-Mills theory might be found in a TNS.

There are, however, still many challenges facing the hypothesis that TNS are useful for the solution of Yang-Mills theory: there is still a large divide between the discrete gauge groups considered in most investigations so far and the compact gauge groups $\su2$ and $\textsl{SU}(3)$ relevant for the standard model (for some progress see \cite{tagliacozzo_tensor_2014, silvi_lattice_2014, zohar:2015, zohar:2016, zohar_projected_2016, kull_classification_2017}). Additionally, there is not yet any systematic way to take a continuum limit of a TNS to obtain a representation of the $n$-point  functions required for a quantum field description. Also, continuum generalisations of the MPS, PEPS, and MERA TNS are available \cite{haegeman:2011a}, but they don't seem particularly well suited for locally gauge-invariant quantum fields.)

In this paper we pursue a description of the ground-state of lattice gauge theory in terms of a TNS. We work with pure gauge theory in the hamiltonian formalism \cite{kogut:1975a} and study the locally gauge invariant sector of hilbert space.  We develop a toolkit to describe states in this sector, exploiting parallel transport operations and block-spin averaging operations to construct hierarchical tensor networks. While the lattice regulator breaks lorentz invariance, time is continuous in the hamiltonian setting; we later argue that lorentz invariance is recovered in the continuum limit as the regulator is removed. This paper is intended as a high-level overview of an ongoing programme: the results reported here are mostly described at a heuristic level and are still replete with conjectures. A more thorough explanation of the results described here is in preparation and will be presented in a series of future papers.

Before we begin, it is worth mentioning that there have been several notable mathematical approaches to the study of Yang-Mills theory, strongly related in spirit to ours. These approaches rely, in various ways, upon the renormalisation group \cite{wilson:1975a} and path-integral type formalisms in terms of action functionals on spacetime. The first programme \cite{balaban:1985a,balaban:1988a,balaban:1984a,balaban:1984b,balaban:1985b,balaban:1985c,balaban:1985d,balaban:1989a,balaban:1989b,balaban:1987a,balaban:1988b}, due to Ba{\l}aban, studies the behaviour of the partition function for lattice gauge theory under the action of block-spin renormalisation operations. This approach has yielded major successes, including, a proof of the ultraviolet stability of the partition function \cite{balaban:1985d} in three spacetime dimensions. Similar to Ba{\l}aban, the second programme \cite{federbush:1987a,federbush:1986a,federbush:1987b,federbush:1987c,federbush:1988a,federbush:1990a}, due to Federbush, yields a continuum limit of the lattice gauge theory as an inductive limit of block-spin renormalisations. The final approach \cite{magnen:1993a} studies pure Yang-Mills in the continuous case, but in the presence of an infrared cutoff. Here the existence of pure Yang-Mills theory is proved and the associated Schwinger functionals constructed. The limit where the cutoff is removed was so far not considered.

The programme outlined here has a different emphasis to the aforementioned mathematical approaches. The ultimate (and perhaps impossible!) objective of this work is to obtain an efficiently contractible representation of the ground state  and low-lying excitations of QCD which will be useful for perturbative and nonperturbative calculations. To do this we are willing to take on faith several key physical assumptions, including, asymptotic freedom and the existence of a spectral gap for lattice gauge theory. Thus we are not so much concerned with questions of existence in the mathematically rigourous sense. That is not to say that the approach here has nothing to say about this question, only that we haven't considered it yet \footnote{If we are to speculate wildly for a moment we would say that an efficiently contractible ansatz for the ground state of Yang-Mills could well lead to progress on question of existence in the constructive QFT sense: one only needs to look to condensed matter physics and the examples of the BCS state and the Laughlin wavefunctions to see how a good ansatz leads to both physical and mathematical progress.}. 

\section{Hilbert space}
We discuss here Yang-Mills theory in the temporal gauge and hamiltonian setting on a regular spatial lattice $a\mathbb{Z}^d$ with lattice spacing $a > 0$ embedded in $\mathbb{R}^d$ \cite{kogut:1975a}. The lattice comprises a set of \emph{directed links} $E$, given by line segments decorated with an arrow, pointing from lattice points $ax\in a\mathbb{Z}^d$ to all neighbouring points $ax+a\widehat{\mu}$, with $\widehat{\mu} \equiv (0, 0, \ldots, 0_{\mu-1}, 1_\mu, 0_{\mu+1}, \ldots, 0)$ and $\mu = 1, 2, \ldots, d$:
\begin{center}
	\includegraphics{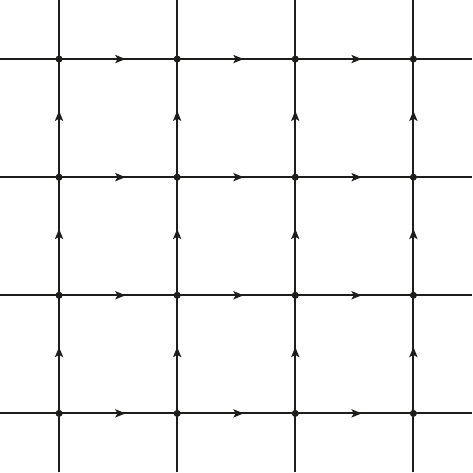}
\end{center}
We sometimes refer to the link pointing from lattice site $ax$ in the direction $\widehat{\mu}$ using the notation $e_{x,\widehat{\mu}}$.

If $e\in E$ is a directed link we denote by $e_-$ the \emph{source} lattice point at the origin of the segment and $e_+$ the \emph{target} lattice point. We often require more general structures than regular lattices: when necessary we work with \emph{directed graphs} $(V,E)$ comprising a set of vertices $V$ and directed edges (or links) $E$. Again we decorate the edges with an arrow to indicate their orientation. 

We attach to each directed link the hilbert space $L^2(G)$ of square integrable functions on a compact lie group $G$. The total hilbert space (of which the \emph{physical} states span a subspace -- see Section~\ref{sec:gis}) is
\begin{equation} \label{eq:total_H}
  \mathcal{H}\equiv \bigotimes_{e\in E} L^2(G).
\end{equation}
Here the two choices $G=\uone$ and $G=\su2$ exemplify the key differences; the more general case does not require the introduction of many new ideas. Note that $\uone$ is diffeomorphic to the circle $S^1$ and $\su2$ is diffeomorphic to the $3$-sphere $S^3$. Thus, the \emph{position} degree of freedom associated with a link is an element of the compact group $G$. Informally the ``position basis'' for such a degree of freedom is written as  
\begin{equation}
	|U\rangle, \quad U\in G,
\end{equation}
with ``inner product''
\begin{equation}
	\langle U|V\rangle = \delta(U,V),
\end{equation}
where $\delta(U,V)$ is the Dirac delta on elements of~$G$. A useful intuitive picture to keep in mind here is that the $\uone$ case is equivalent to a standard Schr\"odinger particle on the circle and the $\su2$ case is equivalent to a particle on the three-dimensional sphere $S^3$.

We often exploit the \emph{left} and \emph{right} rotations $L_U$ and $R_U$ on $L^2(G)$, which are unitary operations given by
\begin{equation}
	L_U|V\rangle \equiv |UV\rangle, \quad \text{and} \quad R_U|V\rangle \equiv |VU^\dag\rangle,
\end{equation} 
where for abelian $G$ we have $R_U = L_U^\dagger$.

We call an assignment of elements of $G$ to the links a \emph{connection} or a \emph{gauge field}. When we have a representation $\pi$ of $G$ we call the matrix representation $\pi(U_e)$ of a link variable $U_e\in G$ a \emph{parallel transporter}. (We often overload this terminology and refer to $U_e$ both as an abstract group element \emph{and} as the defining representation of $G$.)

To a large extent the $\su2$ case subsumes the $\uone$ case, so from now on we frame our discussion in terms of $\su2$. There are some crucial differences, however, which we highlight as we go: as we'll see, in contrast to $\uone$, both the nontrivial curvature and the structure of the representation category of $\su2$ play an important role in the renormalisation of the momentum and kinetic energy operators leading to the spontaneous generation of a gap for $\su2$ Yang-Mills theory.

Formally we work with the hilbert space of square-integrable functions on $G$, whose elements may be represented as
\begin{equation}
	|\psi\rangle = \int dU\, \psi(U)|U\rangle,
\end{equation}
where $dU$ is the Haar measure. The Peter-Weyl theorem shows that $L^2(G)$ may be decomposed as
\begin{equation}\label{eq:peterweyl}
	L^2(G) \cong \bigoplus_{l} V_l\otimes V_l^*,
\end{equation}
where $l$ indexes the irreps, $V_l$ denotes the vector space furnishing the irreducible representation $t^l$, and $V^*_l$ its dual. In the case of $G \cong \uone$ all the $V_l$ are one dimensional and Eq.~(\ref{eq:peterweyl}) reduces to the familiar \emph{fourier decomposition} of functions on the circle. For $\su2$ Eq.~(\ref{eq:peterweyl}) represents a \emph{generalised fourier decomposition} where the ``modes'' have additional structure corresponding to the $V_l$ having dimension $\ge 1$.

The irreducible unitary representations $t^l$ of $\su2$ are labelled by non-negative half integers, $l \in \frac12\mathbb{Z}^+$, and have dimension $d_l = 2l+1$.
According to the Peter-Weyl theorem the matrix elements $t^l_{jk}$ of the irreducible representations furnish a basis for $L^2(\su2)$; we write
\begin{equation}
	|j\rangle_l|k\rangle_l \cong \sqrt{2l+1} t_{jk}^l, \quad l \in \frac12\mathbb{Z}^+,\quad j,k = -l, \ldots, l,
\end{equation}
for the corresponding orthonormal basis, and write the scalar product as
\begin{equation}
	\langle \phi|\psi\rangle = \sum_{l}\sum_{j,k = -l}^l \overline{\widehat{\phi}_{jk}^l}\widehat{\psi}_{jk}^l,
\end{equation}
where 
\begin{equation}
	|\phi\rangle = \sum_{l}\sum_{j,k=-l}^l\widehat{\phi}_{jk}^l |j\rangle_l|k\rangle_l,
\end{equation}
and the summations over $j$ and $k$ are taken in integer steps from $-l$ to $l$. The numbers $\widehat{\phi}_{jk}^l$ are the \emph{fourier coefficients} of $\phi:\su2\rightarrow \mathbb{C}$, and are determined by
\begin{equation}
	\widehat{\phi}_{jk}^l = {_l\langle jk|\phi\rangle} = \sqrt{2l+1}\int dU \, \overline{t^l_{jk}}(U) \phi(U).
\end{equation}

In the case of $\uone$ we may use simpler notation since its irreducible unitary representations are
all one dimensional. They are labelled by the integers $n \in \mathbb{Z}$ so that we may decompose a wavefunction as
\begin{equation}
	|\phi\rangle = \sum_{n=-\infty}^{+\infty} \widehat{\phi}_n |n\rangle,
\end{equation}
with
\begin{equation}
	\widehat{\phi}_n = {\langle n|\phi\rangle} =  \int_0^{2\pi} \frac{d\theta}{2\pi} \, e^{-in\theta} \phi(\theta),
\end{equation}
where we have written out the haar measure in terms of positions on the circle
$0 \le \theta < 2\pi$.

Thanks to Eq.~(\ref{eq:peterweyl}) we know that $L^2(\su2)$ has the structure of a direct sum of bipartite hilbert spaces thus it may be regarded, \emph{for all practical purposes}, as the state space of a bipartite quantum system. We visualise this bipartite-like structure graphically by adding virtual vertices to the ends of a link, intended to represent these two different subsystems: we now picture an arbitrary state $|\psi\rangle$ of a single link as  
\begin{center}
\includegraphics{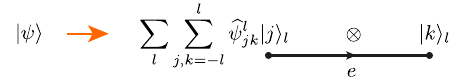} \quad .
\end{center}
Note that we cannot draw this picture in the $\uone$ case because all the irreps of $\uone$ are one dimensional. Nevertheless it is still useful to distinguish the two ends of a link because, as we will see below, they can be manipulated separately using \emph{parallel transport} operations.

According to these observations we extend our original lattice representation by associating the degrees of freedom at the ends of a link with auxiliary vertices associated with the lattice locations:
\begin{center}
	\includegraphics{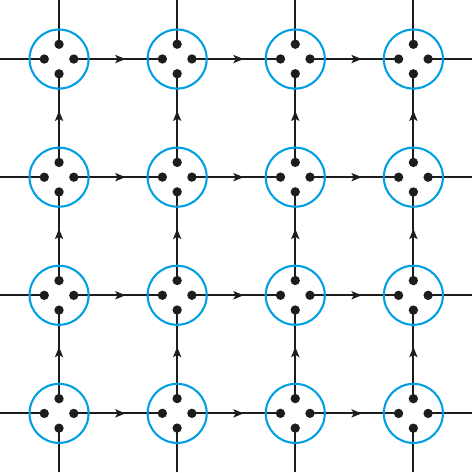}
\end{center}
To distinguish the auxiliary vertices from the lattice sites we refer to the former as \emph{vertices} and the latter as \emph{lattice locations} or \emph{lattice points}.

In two spatial dimensions we have, associated to each lattice location $v$, the tensor product space 
\begin{equation}
	\mathcal{H}_v \equiv \bigoplus_{l_1,l_2,l_3,l_4} V_{l_1}^*\otimes V_{l_2}^*\otimes V_{l_3}\otimes V_{l_4},
\end{equation}
corresponding to the degrees of freedom at the sources and targets of links incident with the corresponding lattice point. The total hilbert space~(\ref{eq:total_H}) can also be written as a product over vertex spaces $\mathcal{H}_v$, together with constraints that fix representation labels to be the same on each link, with a similar decomposition existing for arbitrary graphs \cite{baez:1996a}. However, for our purposes it is sufficient to consider the link decomposition~(\ref{eq:total_H}) while keeping in mind the bipartite structure of links.

Define the $\su2$ \emph{position observables} $\widehat{u}_{jk}$ via
\begin{equation}
	\widehat{u}_{jk}|U\rangle \equiv t_{jk}^{\frac12}(U)|U\rangle \equiv [U]_{jk}|U\rangle,
\end{equation} 
for $j,k \in \{-\frac12, \frac12\}$, i.e., $\widehat{u}_{jk}$ simply gives the matrix elements of the spin-$1/2$ defining representation of $U$ (any other representations could be used here). The $2\times 2$ matrix of operators $\widehat{u}_{jk}$ is simply denoted by $\widehat{u}$. Let $\gamma = (e_1,e_2,\ldots,e_n)$ be a \emph{path} in the lattice, i.e., a sequence of edges $e_j$ such that $e_{j+1,-} = e_{j,+}$. We denote by $\widehat{u}_{jk}(\gamma)$ the \emph{Wilson line} observable
\begin{equation}
	\widehat{u}_{jk}(\gamma) \equiv \widehat{u}_{j j_1}(e_1)\widehat{u}_{j_1j_2}(e_2)\cdots \widehat{u}_{j_{n-1} k}(e_n),
\end{equation}
where $\widehat{u}(e)$ denotes the position observable for link $e$ and repeated indices are summed. Note that $\widehat{u}_{jk}(e) = \widehat{u}_{jk}$ in the case $\gamma$ traverses edge $e$ \emph{against} the direction of $e$ and $\widehat{u}_{jk}(e) = \widehat{u}^{\dagger}_{kj}$ if the traversal direction is \emph{the same} as the direction of $e$. In the case where the path $\gamma$ is closed we set $k=j$ and sum over the index $j$; such observables are denoted 
\begin{equation}
	\tr(\widehat{u}_{\gamma}) \equiv \sum_{j}\widehat{u}_{j j_1}(e_1)\widehat{u}_{j_1j_2}(e_2)\cdots \widehat{u}_{j_{n-1} j}(e_n),
\end{equation}
and are known as \emph{Wilson loops}. We explicitly allow the case where a path visits a given link/edge more than once. In the special case where the path $\gamma$ traverses a \emph{plaquette} or \emph{face} in the clockwise direction we denote the Wilson loop as $\tr(\widehat{u}_{\cwplaqsub})$.

For $G \cong \uone$ the position observable
\begin{equation}
	\widehat{u}|\theta\rangle \equiv e^{i\theta}|U\rangle
\end{equation}
has no additional indices
so that the trace is not needed when defining loops: 
$\tr(\widehat{u}_{\cwplaqsub}) = \widehat{u}_{\cwplaqsub}$.

Consider now the Lie algebra $\lasu2$ of \su2:
\begin{equation}
	[\tau^j, \tau^k] = -2{\varepsilon^{jk}}_l\tau^l.
\end{equation}
These generators, which play the role of momentum operators, are represented on $L^2(G)$ via infinitesimal rotations by $e^{\epsilon \tau^\alpha}$:
\begin{equation}
	\widehat{\ell}^\alpha_{L} \equiv \frac{d}{d\epsilon} L_{e^{\epsilon \tau^\alpha}},
\end{equation}
with a similar definition for $\widehat{\ell}^\alpha_{R}$. The commutation relations between the position and momentum observables is
\begin{equation}
	[\widehat{\ell}_L, \widehat{u}_{jk}]= -[\tau^\alpha \widehat{u}]_{jk}.
\end{equation} 
In terms of the momentum operators we have the casimir element, the \emph{laplacian}
\begin{equation}
	\triangle = \sum_{\alpha = 1}^3 (\widehat{\ell}_L^{\alpha})^2 =  \sum_{\alpha = 1}^3 (\widehat{\ell}_R^{\alpha})^2 = \bigoplus_{l} (d_l^2-1) \mathbb{I}_l.
\end{equation}

\section{The Kogut-Susskind hamiltonian}
The model we study in this paper arises as the spatial discretisation of the Yang-Mills lagrangian in the temporal gauge. A natural hamiltonian generating the dynamics of pure gauge theory in this setting is that of Kogut and Susskind
\cite{kogut:1975a}:
\begin{equation}
	H(g_H) = -\frac{g_H^2}{2 a} \sum_{e\in E} \triangle_e + \frac{1}{g_H^2 a}\left(2- \sum_{\square} \text{Re}(\tr(\widehat{u}_{\cwplaqsub}))\right),
\end{equation}
where the sum is over all plaquettes of the lattice graph.  

In many ways this is a remarkably simple model: it comprises only two really different terms, a ``kinetic energy'' term $\text{KE} \equiv -\sum_{e\in E} \triangle_e$ and a ``potential energy'' term $\text{PE} \equiv \sum_{\square} 2 -\text{Re}(\tr(\widehat{u}_{\cwplaqsub}))$, both of which can be exactly diagonalised. The $\text{KE}$ term is diagonal in the $|jk\rangle_l$ ``momentum'' basis and the $\text{PE}$ term is diagonal in conjugate ``position'' basis $|U\rangle$, $U\in G$.

The physics of this model is therefore determined by the competition between the $\text{KE}$ and $\text{PE}$ terms: the \emph{lattice strong-coupling} ($g_H\rightarrow \infty$) limit ground state is a product state $|\Omega(\infty)\rangle = \bigotimes_{e\in E} |00\rangle_0$ and the \emph{lattice zero-coupling} ($g_H=0$) limit ground state is a superposition of all \emph{flat} gauge-field configurations in the position basis, namely, those connections satisfying $\prod_{e\in \square} U_e = \mathbb{I}$ for every plaquette $\square$. Note that for both of these extremes this lattice model gives \emph{no} nontrivial dynamics: this is why we insisted on calling the two extremes the ``lattice" strong- and zero-coupling limits in order to distinguish them from the \emph{field} strong and zero-coupling limits (discussed below) which \emph{do} yield nontrivial dynamics.

Notice that the lattice length $a$ plays no role in the diagonalisation of $H$, except to set the overall energy scaling; in pure Yang-Mills theory the ground state acquires a lengthscale only through the spontaneous generation of a spectral gap:  by matching the energy $\Delta E(g_H)$ of the first excited state to the mass of the observed fundamental excitation \footnote{The (conjectured) fundamental excitation of pure Yang-Mills theory, the \emph{glueball}, has not yet been observed directly.}, we determine $a$ as a function of $g_H$. Alternatively, because of the gap, the ground state $|\Omega(g_H)\rangle$ of $H(g_H)$ will have a correlation function decaying exponentially with separation according to a dynamically generated correlation length $\xi(g_H)$. Thus we can also fix $a$ by demanding that $a\xi(g_H)$ tends to a constant. This phenomena is known as \emph{dimensional transmutation} and is a familiar feature of condensed matter systems in the \emph{scaling limit} \cite{sachdev:2011a}.

The way a lattice system such as the Kogut-Susskind model describes a continuous quantum field is via a \emph{scaling limit} of a second- or higher-order quantum phase transition: we need to find a point where the correlation length $\xi(g_H)$ (measured in lattice sites) diverges so that the corresponding lattice spacing --- fixed by $a\xi(g_H) = \text{const.}$ --- goes to $0$. 
Our task thus becomes to locate the second-order phase transitions for the KG model and to analyse their scaling limits. 

It is now understood that \emph{nonabelian} Yang-Mills theory is \emph{asymptotically free}. What this means in the present context is that the spectral gap $\Delta E(g_H)$ between the ground and first-excited states of $H(g_H)$ is nonvanishing for all $g_H >0$ and only disappears when $g_H$ is exactly zero \footnote{Presumably $\Delta(g_H)$ could be monotonically decreasing as a function of $g_H$, but this is unlikely as the addition of irrelevant ultraviolet interactions could easily modify the behaviour of the gap for large coupling $g_H$ without leading to any impact on the large-scale physics.}. In the condensed-matter context we understand asymptotic freedom as saying that the only quantum phase transition for the model occurs at \emph{exactly} $g_H=0$ and that this transition is (at least) second order; this is a property shared by, e.g., rotor models \cite{sachdev:2011a}. Asymptotic freedom has profound consequences for the analysis of Yang-Mills theory as it allows us to apply perturbation theory around the exactly solvable field zero-coupling point. In this way many deep insights have been obtained into high energy processes. It is worth pointing out that asymptotic freedom for the lattice gauge theory is not universally accepted \cite{seiler:2003a}: it is logically possible that the KG model has a phase transition at some intermediate value of $g_H$ and that pure Yang-Mills theory corresponds somehow to the scaling limit of that transition. 

We proceed under the conventional assumption that $\su2$ lattice gauge theory is asymptotically free. (As we later explain, it is actually not inconceivable that the ansatz described here might lead to a proof of asymptotic freedom for the $\su2$ Kogut-Susskind hamiltonian.) We exploit this property to guarantee that our ground-state ansatz for the model applies throughout the entire range of values of $g_H$ from $\infty$ down to $0$.

The Kogut-Susskind hamiltonian is a regulated version of quantum Yang-Mills theory. Thus, presumably, we can express
\begin{equation}
	H_{\text{KS}}(g_H) = H_{\text{YM}}(g) + H_{\text{C}}(\Lambda),
\end{equation}
where $H_{\text{YM}}(g)$ is the full nonperturbative hamiltonian for continuum quantum Yang-Mills theory with some definite value of the \emph{Yang-Mills coupling constant} $g$ and $H_{\text{C}}(\Lambda)$ is a regulator with cutoff $\Lambda \equiv \Lambda(g)$. There is a complicated relationship between $\Lambda$, $g_H$, and $g$ determined by the renormalisation group. In a condensed-matter physics language, these statements are equivalent to saying that $H_{\text{KS}}$ is the same as a \emph{critical model} $H_{\text{YM}}(g)$ in the presence of an external field $H_{\text{C}}(\Lambda)$. Our objective is then, given only $H_{\text{KS}}(g_H)$, to obtain $H_{\text{YM}}(g)$ by removing the cutoff $H_{\text{C}}(\Lambda)$. This amounts to studying the quantum phase transition at $g_H = 0$.

Note that the question of whether the Kogut-Susskind hamiltonian $H_{\text{KS}}(g_H)$ has a gap $\Delta E(g_H)$ for all $g_H > 0$ is logically distinct from the (considerably more challenging) problem of determining whether the full nonperturbative \emph{quantum Yang-Mills theory} $H_{\text{YM}}(g)$ has a spectral gap, a question which is (part of) the content of the first Clay maths millennium problem \cite{cmproblem:1st}. If the programme set out in this paper were to come to fruition then we should be able to deduce that $\Delta E(g_H) >0$ implies the \emph{existence} of quantum Yang-Mills theory. Whether our ground-state ansatz could ever lead to a proof of the mass gap is somewhat more tendentious but, hopefully, not impossible.

It is now worth contrasting the $\su2$ case with the $\uone$ case: it turns out that \emph{abelian} Yang-Mills theory on the lattice, known as \emph{periodic} or \emph{compact} Maxwell theory, is dramatically different \cite{polyakov:1975a, villain:1975a, polyakov:1977a, banks:1977a, peskin:1978a, kogut:2004a} to its nonabelian counterpart. It is understood that, in the abelian case, there is a phase transition --- possibly of second order --- at a finite value of $g_H$. This means that the strong-coupling phase is separated from the zero-coupling phase and the scaling limit found from approaching the critical value of $g_H$ from the strong-coupling side is a physically distinct quantum field theory --- known to be confining --- from the scaling limit found by approaching the transition from the zero-coupling side, which, presumably, gives us standard $\uone$ gauge theory. In order to approximate standard $\uone$ gauge theory we would need to start from the zero-coupling fixed point and develop an ansatz which approaches the phase transition from below. As will become evident, our ansatz is not well-suited for this task and, in the $\uone$ case, is likely to only describe the physics of the strong-coupling phase. 

The Kogut-Susskind hamiltonian still possesses a tremendous amount of local symmetry: although we are working in the temporal gauge any constant gauge transformation is still a symmetry of the model. We elaborate on this local gauge symmetry in Sec.~\ref{sec:gis}.

\section{The problem}\label{sec:problem}
We have now collected enough preliminary material to finally describe the problem we are aiming to solve. We want to find a one-parameter family of quantum states $|\Phi(g_H)\rangle$ for hamiltonian lattice gauge theory with the following properties:
\begin{enumerate}
	\item The state $|\Phi(g_H)\rangle$ is an efficiently contractible tensor network state for all dimensions $d$ and all couplings $g_H$.
	\item It interpolates between the lattice zero-coupling and lattice strong-coupling limits where, respectively, $|\Phi(0)\rangle = |\Omega(0)\rangle$ and $|\Phi(\infty)\rangle = |\Omega(\infty)\rangle$. 
	\item The TNS $|\Phi(g_H)\rangle$ is (manifestly) locally gauge invariant.
	\item The state $|\Phi(g_H)\rangle$ differs from $|\Omega(g_H)\rangle$ only by \emph{irrelevant} UV features. That is, $|\Phi(g_H)\rangle$ is the ground state of a \emph{parent} hamiltonian $H'(g_H)$ such that the operator $H'(g_H)-H(g_H)$ is irrelevant for the RG. 
	\item The continuum limit of $|\Phi(g_H)\rangle$ may be analytically obtained and is lorentz invariant.
\end{enumerate}

\section{The gauge invariant sector}\label{sec:gis}
Although we are working in the temporal gauge there is still a residual gauge freedom \cite{creutz:1985a}. This freedom is respectively expressed in terms of the \emph{gauge group} $\mathcal{G}$ which is the cartesian product of the group $G \cong \su2$ or $\uone$ over all the lattice points, $\mathcal{G}\cong \prod_{v\in V} G$, and which is represented on $\mathcal{H}$ by 
\begin{equation}
	x \mapsto \bigotimes_{e\in E} L_{x_{e_-}}R_{x_{e_+}}, \quad x\in \mathcal{G}.
\end{equation}
All physical states live in the \emph{gauge-invariant subspace} $\mathcal{H}_{\mathcal{G}}$ of $\mathcal{H}$, which is the subspace  spanned by all vectors satisfying
\begin{equation}
	\bigotimes_{e\in E} L_{x_{e_-}}R_{x_{e_+}}|\psi\rangle = |\psi\rangle, \quad \forall x \in \mathcal{G}.
\end{equation}

The most important gauge-invariant state is built from the trivial representation of $G$ and is given by
\begin{equation}
	|\omega_0\rangle = \int |U\rangle\, dU.
\end{equation}
The wavefunction for this state is simply the constant function corresponding to the basis vector, $|00\rangle_0 \cong t^0_{00}(g)$ for $\su2$ and $|0\rangle \cong e^{i0\theta} = 1$ for $\uone$. 
Note that left and right invariance of the haar measure implies that
\begin{equation}
	L_U|\omega_0\rangle = R_U|\omega_0\rangle = |\omega_0\rangle, \quad \forall U\in G.
\end{equation}
Using $|\omega_0\rangle$ we can build the state 
\begin{equation}
	|\Omega_\infty\rangle = \bigotimes_{e\in E} |\omega_0\rangle,
\end{equation}
which is gauge invariant with respect to an \emph{arbitrary} graph.

In the next section we show how to represent any state in the gauge-invariant sector as a TNS. Before moving on to this we briefly cover the special case of lattice gauge theory on a single lattice point with a single edge, i.e., a single loop:
\begin{center}
	\includegraphics{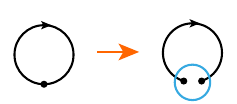}
\end{center}
The hilbert space for this system is $L^2(G)$ and the gauge-invariant sector $\mathcal{H}_{\mathcal{G}}$ is found as follows. The local gauge group $\mathcal{G}$ acts as
\begin{equation}
	|\psi\rangle \mapsto L_xR_x |\psi\rangle, \quad \forall x\in G.
\end{equation}
Thus a state $|\psi\rangle$ is gauge invariant if and only if
\begin{equation}\label{eq:classstate}
	\int \psi(U) |U\rangle\, dU = \int  \psi(x^{-1}Ux) |U\rangle\, dU, \quad\forall x\in G.
\end{equation}
Because $\mathcal{G}$ acts \emph{irreducibly} $\psi$ must be a \emph{class function}, i.e., $\psi(x^{-1}Ux) = \psi(U)$. Note that this is a trivial requirement in the abelian case $G \cong \uone$, where we have $L_xR_x = \mathbb{I}$.

A generalisation of the single-loop graph is the \emph{petal} graph 
\begin{center}
	\includegraphics{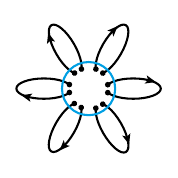}
\end{center}
comprising a single vertex and $n$ edges. The configuration space for this system is $\mathcal{H} = \bigotimes_{j=1}^n L^2(G)$. The local gauge group $\mathcal{G}$ is again isomorphic to a single copy of $G$ and now acts as a tensor product:
\begin{equation} \label{eq:petalinvariance}
	|\psi\rangle \mapsto (L_xR_x)^{\otimes n} |\psi\rangle, \quad \forall x\in \mathcal{G},
\end{equation}
i.e., a gauge-invariant wavefunction obeys $\psi(x^{-1}U_1x, x^{-1}U_2x, \ldots, x^{-1}U_nx) = \psi(U_1, U_2, \ldots, U_n)$, $\forall x\in \mathcal{G}$.
Since $\mathcal{G}$ now acts \emph{reducibly} the gauge-invariant sector is no longer spanned by states whose wavefunctions are class functions; instead a gauge invariant state for the petal graph is equivalent to a sum of superpositions of \emph{intertwiners} \cite{baez:1996a}, with the sum running over every sector of $\mathcal{H}$ obtained by fixing the irrep of $G$ in~(\ref{eq:peterweyl}) for each petal.

As we'll see, the single-loop and petal graph cases turn out to play an important role in the construction of general gauge-invariant states for arbitrary lattices.

\section{Parallel transport}\label{sec:paralleltx}
The classical notion of parallel transport through a gauge field on a lattice is described as follows. Suppose that we have an object transforming according to a representation $\pi$ of $G$. We think of the object as living at some vertex $v$. Whenever the object moves to another vertex $w$ along a path $\gamma$ it undergoes the \emph{parallel transport}  
\begin{equation}\label{eq:ugamma}
	\pi(\gamma) \equiv \prod_{e\in \gamma} \pi(U_{e}^{-\sgn_\gamma{e}}),
\end{equation}
where $\sgn$ is $+1$ or $-1$ according to whether the link is traversed in the direction of the arrow or not and the product is taken from \emph{right to left}.

The quantum representation of the parallel transport is furnished by a fundamental primitive called the \emph{controlled rotation} unitary gate
\begin{equation}
	\CU \equiv \int |U\rangle\langle U| \otimes \pi(U)\,dU,
\end{equation}
where $\pi$ is unitary representation of $G$ on a vector space $V_\pi$. The controlled rotation is a unitary operator on $L^2(G)\otimes V_\pi$; the first tensor factor is called the \emph{control} and the second factor the \emph{target}. We think of $V_\pi$ as the configuration space of a quantum particle initially located at the source of the link and, after the application of $\CU$ the particle has been transported to the target of the link. To describe more complicated parallel transport operations let $\gamma$ be a path in $(V,E)$ and denote by
\begin{equation}
	\begin{split}
		\CU_\gamma &\equiv \int   \left(\bigotimes_{e\in \gamma} |U_e\rangle\langle U_e|\right) \otimes \pi(\gamma)\, \left(\prod_{e\in \gamma} dU_e\right) \\
		&= \prod_{e\in \gamma} {\CU_{e s}}^{-\sgn_\gamma{e}},
	\end{split}
\end{equation}
where the product is taken, as usual, from right to left, and $\pi(\gamma)$ is defined as in (\ref{eq:ugamma}). It is a simple calculation to deduce that the transported particle transforms in the correct way under the gauge group after the transport operation.

It is clear that $\CU_\gamma$ is an \emph{entangling} operation and, hence, there is no way in general to separate the gauge degrees of freedom from a quantum particle's position degree of freedom after it has undergone parallel transport: these two degrees of freedom typically become strongly entangled during parallel transport.

Thus far in our discussion here the target of the parallel transport gate has been an \emph{additional} degree of freedom so that the appropriate hilbert space for this system is $\mathcal{H}\otimes V_\pi$. In this way it is a simple matter to introduce additional fields at the vertices, e.g.\ fermions, which correspond to quarks, or bosons, appropriate for Higgs models (this will be investigated in a later paper). However, our focus here is on \emph{pure} Yang-Mills theory, so we now describe how to exploit the quantum parallel transport operation directly without the introduction of ancillary degrees of freedom.

In pure gauge theory a vertex at the end of a link may itself be regarded as a quantum particle transforming according to a representation of $G$, namely, the left or right regular representation. Hence, we can exploit parallel transport to move these vertices (and their associated edges) around the gauge network. This operation is effected by using for the representation $\pi(U)$ either the left and right multiplication operations $L_U$ or $R_U$ as follows. Suppose we wish to move the target vertex $v = f_+$ of an edge $f \in E$ to some other lattice point $w$ along a path $\gamma$. Then we simply apply the operation
\begin{equation}
	\CR_\gamma \equiv \prod_{e\in \gamma} {\CR_{e f}}^{-\sgn_\gamma{e}}
\end{equation}
Note that planarity of the graph $(V,E)$ is not relevant for this operation: the procedure is identical for any oriented graph.

By combining parallel transport with gauge-invariant loops we can describe an important primitive, namely, \emph{edge addition}. Suppose we have a gauge theory on a graph $(V,E)$ in a gauge-invariant state $|\Phi\rangle$. We produce a new state $|\Phi'\rangle$ for a graph $(V, E\cup (v,w))$ with an additional edge $(v,w)$  between any two lattice locations $v$ and $w$ via the following sequence of steps. \begin{enumerate}
	\item The first step is to add in a loop in a state $|\psi\rangle$ in the gauge-invariant sector based at the source $v$ of the new edge:
	\begin{equation}
		|\Phi\rangle \mapsto |\Phi\rangle |\psi\rangle.
	\end{equation}
	\item Now parallel transport the target vertex of the loop to its destination lattice location $w$ along a path $\gamma$:
	\begin{equation}
		|\Phi\rangle |\psi\rangle \mapsto \CR_\gamma |\Phi\rangle |\psi\rangle.
	\end{equation}
\end{enumerate}
The resulting state $|\Phi'\rangle \equiv \CR_\gamma |\Phi\rangle |\psi\rangle$ is a state of a lattice with an additional edge $e$ between $v$ and $w$. Further, it transforms correctly under the gauge group $\mathcal{G}$. Note: unless the original state $|\Phi\rangle$ of the lattice is \emph{flat} (more on this later) the result of this operation generally depends on the path $\gamma$ chosen between $v$ and $w$. 

To understand how to produce all vectors within the gauge-invariant sector we need an additional operation, namely, \emph{lattice point addition} or \emph{edge subdivision}. Suppose that $|\Phi\rangle$ is a gauge-invariant state for a graph $(V,E)$ and we wish to subdivide an edge $e = (v,w)$ by adding a lattice point: we want to obtain a new gauge-invariant state for the graph $(V',E')$ where $V' = V\cup \{v'\}$ and $E' = (E\setminus \{e\} )\cup\{ (v,v'), (v',w)\}$. This is carried out using the procedure:
\begin{enumerate}
	\item Adjoin an ancillary subsystem in the state $|\omega_0\rangle_{e'}$, where $e' = (v,v')$, resulting in the new gauge-invariant state
	\begin{equation}
		|\Phi\rangle |\omega_0\rangle_{e'}.
	\end{equation} 
	(The reason this state is gauge invariant on the bigger graph with vertex set $V\cup \{v'\}$ is because of the separate left- and right-invariance of the haar measure.)
	\item Apply $\CL^{-1}$ to glue the new edge to the end of the old edge $e$:
	\begin{equation}
		\CL_{e' e}^{-1}|\Phi\rangle |\omega_0\rangle_{e'}.
	\end{equation}
	\item Relabel the subsystem $e$ as $e'' = (v',w)$. We end up with the state
	\begin{equation}
		|\Phi'\rangle = \int d\mathbf{U}dU_{e'}dU_{e''} \, \Phi(\mathbf{U},U_{e''}) |\mathbf{U}\rangle |U_{e'}\rangle_{e'}  |U^{\dag}_{e'}U_{e''}\rangle_{e''},
	\end{equation}
\end{enumerate}
where $\mathbf{U}$ refers to the connection variables attached to edges in $E\setminus \{e\}$.
The edge subdivision procedure is simply a parallel transport of the source vertex of $e$ along a new edge $e'$ initialised in the trivial state $|\omega_0\rangle$.

\begin{figure*}
	\includegraphics{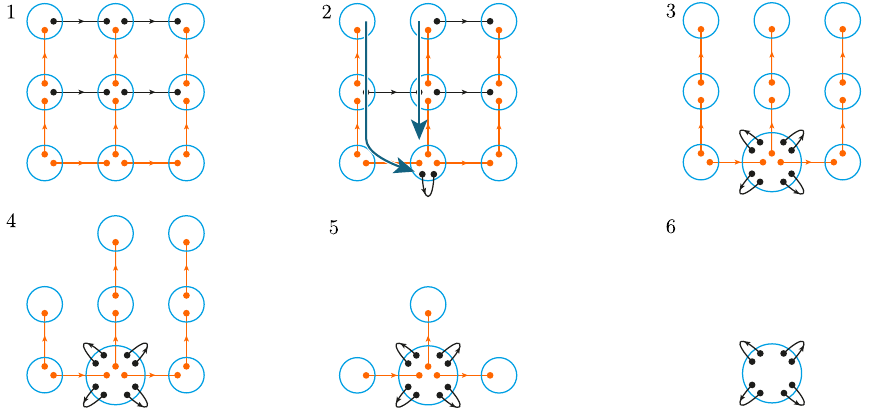}
	\caption{An illustration of the procedure to identify the gauge-invariant sector $\mathcal{H}_{\mathcal{G}}$ for a $3\times 3$ grid. First a maximal tree is found (indicated with the orange edges). Then the remaining edges are parallel transported to the root vertex. Finally, the maximal tree is contracted to the origin.}\label{fig:gis}
\end{figure*}

We now have all the ingredients to describe how to construct an arbitrary state in the gauge-invariant sector. The key observation is that both the operations of edge and lattice point addition are isometric and can be applied \emph{in reverse} to move edges to loops at a single lattice location and, crucially, to isometrically \emph{remove} lattice points with degree $2$. The following argument is reminiscent of the proof of the van Kampen theorem \cite{hatcher:2002}.

Suppose we have an arbitrary state $|\Phi\rangle$ of a graph $(V,E)$. The first step is to identify a \emph{maximal tree}, which is a maximal connected subgraph $T$ of $(V,E)$ containing no loops (there are many such subgraphs of any graph). Note that all the vertices $V$ of our initial graph participate in $T$. Locate and mark the root lattice site $v$ of the tree. All the edges which aren't in $T$ have their ends on lattice points in the tree. The next step is to parallel transport all of these edges along the tree to loops based at the root vertex $v$. We are now left with a bunch of loops based at $v$ and the tree $T$. The final step is to apply lattice point removal to contract all the leaves of $T$ to the root site $v$. At the end of this process we end up with a petal graph of loops based at a single lattice point $v$. 
Since this entire process was isometric we see that the gauge-invariant subspace of $\mathcal{H}$ is equivalent to the space of gauge invariant states of a petal graph, which, in turn is given by the space of (superpositions of) equivariant maps. An illustration of this procedure is shown in Fig.~\ref{fig:gis}. Note that, for $G \cong \uone$ the gauge-invariant space of a petal graph is the full Hilbert space of the petal graph; there is \emph{no residual gauge freedom} because Eq.~(\ref{eq:petalinvariance}) is trivial, unlike for $G \cong \su2$.

Notice that we never needed to fix a gauge to carry out this process. In lattice gauge theory there is an entirely analogous procedure used to \emph{fix a gauge}: a maximal tree is identified and the gauge freedom is then fixed on the tree. For a discussion of this in the hamiltonian formalism see \cite{ligterink:2000a}. In contrast, we describe an \emph{isometry} that bijectively maps gauge-invariant states of the initial graph to those of the corresponding petal graph. There is no explicit fixing of degrees of freedom. Rather, it is a consequence of the parallel-transport map that the edges on the maximal tree are left in a particular (product) state, such that they can be discarded.

Using the primitives of edge and lattice point addition described here we can construct arbitrary gauge invariant tensor network states. There are two equivalent ways to do this: either (a) first parallel-transport everything to a single lattice point and apply a standard ($\su2$ invariant) tensor network ansatz to the remaining degrees of freedom; or (b) generalise the description of  tensor networks as contractions over \emph{ancillary degrees of freedom} by instead using parallel transport operations to introduce the auxiliary degrees of freedom in an explicitly gauge-invariant way. The former approach suffers slightly from the fact that the Kogut-Susskind hamiltonian looks somewhat nonlocal in terms of the remaining degrees of freedom. Thus we exploit the latter approach in the sequel. Note that building gauge-invariant tensor networks for fermions is a straightforward generalisation of the pure gauge case described here. 

\section{The renormalisation group for lattice gauge theory: interpolation}
In this and the following sections we describe the renormalisation group for lattice gauge theories. We define a block-spin type RG which has both the strong coupling and zero coupling limit ground states as exact fixed points.  By finding an inverse to this RG we are able to propose a ground-state ansatz interpolating between the two fixed points. We begin our development with a discussion of the ground-state physics of the Kogut-Susskind hamiltonian which we then use as inspiration for a procedure to interpolate between the two limiting cases.

There is a natural competition between the two terms in the Kogut-Susskind hamiltonian: the kinetic energy term, diagonal in the momentum basis, wants to disentangle each edge and put it into the trivial representation and the potential energy term, diagonal in the position basis, wants to put the lattice into a flat configuration. This competition between momentum and position is a familiar situation from the perspective of nonrelativistic quantum theory.

Our ansatz takes direct inspiration from the intimate connection between the coupling constant $g_H$ and \emph{scale change}.  Let's start with what we know to be true, namely, that the ground state at strong coupling $g_H\rightarrow \infty$ is given by $|\Omega_\infty\rangle$. This state is just about as far from a superposition of flat configurations as possible. The correlation length of this state is, very roughly speaking,  $1$ lattice spacing.  Now imagine changing $1/g_H$ from $0$ to $\epsilon$. What this does is introduce correlations by building small clusters of nearly flat gauge connections. Thus the correlation length increases and we \emph{rescale} $a$ to compensate (recall that the correlation length is a fixed physically observable quantity). Thus a decrease in $g_H$ corresponds to a \emph{scale changing} operation: we are ``zooming into'' the lattice. In any theory with a cutoff a scale-changing operation always brings in new degrees of freedom \cite{haegeman:2011a} --- we have to come up with a way to assign a quantum state to these new degrees of freedom.

In the processing of digital images there is a well-understood prescription to zoom into an image, namely, \emph{interpolation}. We exploit this idea by developing a gauge-invariant interpolation algorithm for (nonabelian) lattice gauge theories to successively build larger and larger clusters. The interpolation method we describe here is directly motivated by \emph{laplace interpolation} for scalar fields $\phi$ (the intensity field). This works by interpolating missing pixels in an image by minimising the (lattice) laplace operator $\triangle \phi$ subject to the boundary conditions supplied by the existing pixels. This method does have some defects due to the singularities of the laplacian green function; we'll explain later how to overcome this. 

There is a pleasing congruence between the notion of interpolation described here and the procedure of \emph{curvature minimisation}: suppose we want to introduce new link degrees of freedom in \emph{as flat a way as possible}. The best way to achieve this would be to maximise the curvature $\text{Re}\tr(\widehat{u}_{\cwplaqsub})$ operator over the new degrees of freedom subject to the constraint that the parallel transporter between the old lattice points remain unchanged. But the lattice curvature operator is the most natural analogue of the spatial laplacian for lattice gauge fields, so we are just doing laplace interpolation. An additional bonus is that this procedure is \emph{gauge invariant}, so we always remain in the gauge-invariant sector.

\subsection{Interpolation for classical nonabelian gauge fields}
In this subsection we describe our interpolation procedure, as applied to \emph{classical} gauge-field configurations. In the next subsection we describe how to use this method to obtain a quantum interpolation prescription. We also note that interpolation of classical lattice-gauge configurations is potentially of broader interest. Indeed, an approximate interpolation procedure has recently been used to improve the performance of Monte Carlo simulations of Yang-Mills theory \cite{endres:2015}.

As a warm up we first consider the problem of interpolation for a tiny ``lattice'' of two points and two edges with $G \cong \su2$. Here the task is to interpolate between the parallel transporters, $U_0$ and $U_n$, on the two edges:
\begin{center}
	\includegraphics{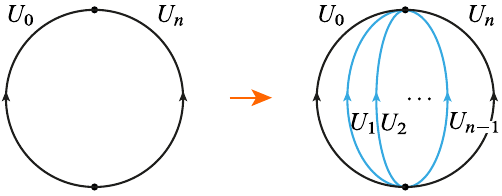}
\end{center}
The solution may be  found variationally. Consider the expression for the total curvature of the subdivided plaquette:
\begin{equation} \label{eq:interpmin}
 4n- 2 \sum_{j=0}^{n-1}  \text{Re}\tr(U_j^\dag U_{j+1}) 
	=  \sum_{j=0}^{n-1} \|U_j-U_{j+1}\|_2^2 .
\end{equation}
We minimise this quantity over ${U_1, \ldots, U_{n-1} \in \su2}$ and find the variational optimum is achieved by
\begin{equation} \label{eq:slerp}
	U_j = U_0 (U_0^\dag U_{n})^{\frac{j}{n}}, \quad j = 1, 2, \ldots, n-1,
\end{equation}
where for the minimum of Eq. (\ref{eq:interpmin}) we take the principal $n$th root.
Notice that this solution transforms correctly under local gauge transformations. Incidentally, this interpolation procedure is familiar from 3D animation and robotics where it is known as SLERP \cite{shoemake:1985a}. 
Note that, since $\uone \subset \su2$, the same result applies in case $G \cong U(1)$ with $U_j = e^{i\theta_j} \in \uone$. 

Motivated by this simple case we now tackle the general interpolation problem: suppose we have a plaquette with $n$ edges. We study the task of subdividing the plaquette and the corresponding parallel transporters $U_e$ into subplaquettes in as ``flat'' a way as possible:
\begin{center}
	\includegraphics{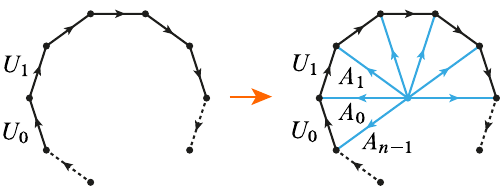}
\end{center}
To this end we study the problem of minimising, over $A_j\in \su2$, $j = 0, 1, \ldots, n-1$, the curvature of the subdivided plaquette:
\begin{equation}
	E=4n-2\sum_{j=0}^{n-1} \text{Re}(\tr(U_j A_j^\dag A_{j-1})),
\end{equation}
where $U_j\in \su2$, $j=0, 1, \ldots, n-1$. Some lengthy derivations supply us with  $n$ possible extremal solutions to our variational interpolation problem, namely,
\begin{equation}
	A_j =   \mu^{-jk}\Phi^{\frac{j}{n}}\eta^\dag U_0 \cdots U_{j}, \quad k = 0, 1, \ldots, n-1,
\end{equation}
where
\begin{equation}
	\eta^\dag U_{n-1}^\dag\cdots U_0^\dag \eta = \begin{pmatrix} e^{i\phi} & 0 \\ 0 &  e^{-i\phi}\end{pmatrix}\equiv \Phi
\end{equation}
and $\mu^{jk} = \left(\begin{smallmatrix} e^{\frac{2\pi i jk}{n}} & 0 \\ 0 & e^{\frac{-2\pi i jk}{n}}\end{smallmatrix}\right)$.
The interpolated curvature for these solutions becomes
\begin{equation}
	E = 4n-4n \cos\left(\frac{\phi - 2\pi k}{n}\right).
\end{equation}
Depending on the \emph{flux} $\Phi$ through the original plaquette it may be necessary to take $k\not=0$ in order to achieve the variational optimum. If $\phi$ is close to zero, however, we see that the total interpolated curvature of the subdivided plaquette scales as $\phi^2/n$. Thus we see that the flux $\Phi$ per plaquette undergoes the transformation $\Phi\mapsto \Phi^{\frac{1}{n}}$ under interpolation, i.e., the flux is simply divided into $n$ pieces and redistributed equally amongst the $n$ new plaquettes.

Checking for consistency with the solution of Eq.~(\ref{eq:interpmin}) we find that, for $n=2$, we recover (for $k=0$) $A_1^\dag A_0 = U_0(U_0^\dag U_1^\dag)^{\frac{1}{2}}$ as expected from Eq. (\ref{eq:slerp}). As before, the above solutions also apply when $G \cong \uone$, in which case $\eta = 1$, $\Phi = e^{i\phi} = U_{n-1}^\dag \dots U_0^\dag$, and $\mu^{jk} = e^{\frac{i2\pi jk}{n}}$. 

Suppose we have a regular lattice in two spatial dimensions. If we successively subdivide plaquettes $m$ times then the curvature \emph{per plaquette} in the refined lattice scales as $\phi^2/4^m$.  If we don't rescale lattice spacing this means that, as $m\rightarrow \infty$, the interpolated connection tends exponentially quickly to a flat connection.

The aforementioned interpolation operations can be generalised to obtain a classical interpolation algorithm applicable for lattices of arbitrary spatial dimension and structure. The cases of central physical interest are $d=1, 2, 3$. In the one dimensional case of a lattice gauge theory on a cylinder we can use the first method to interpolate that gauge field. In the two-dimensional case we exploit the second method as follows. We first subdivide the edges arbitrarily by replacing a connection variable $U$ with a pair $(UX, X^\dag)$, with $X$ arbitary (it makes no nontrivial contribution to gauge-invariant observables). We then apply the interpolation procedure to the additional four vertices:
\begin{center}
	\includegraphics{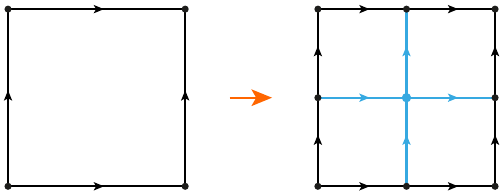}
\end{center}

The three-dimensional interpolation operation is carried out in three steps. Firstly the edges are subdivided as before. Then the two-dimensional algorithm is applied to multiple times to interpolate the faces of each cube:
\begin{center}
	\includegraphics{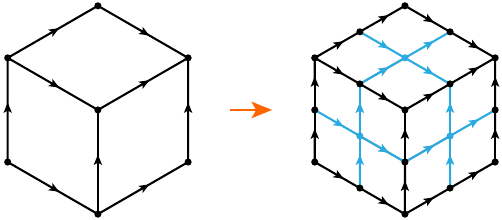}
\end{center}
The interior is then finally subdivided into 8 subcubes using a generalisation of the two-dimensional procedure:
\begin{center}
	\includegraphics{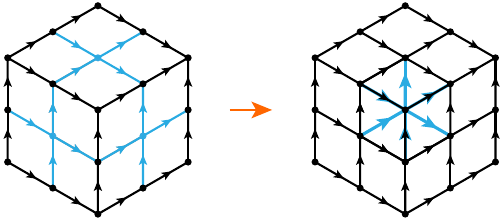}
\end{center}
We detail the steps of this calculation in a future paper.

In general, apart from special cases detailed here, when we want to interpolate a general graph in an arbitrary way we encounter a nonlinear eigenvalue problem. Even though this problem does not, in general, admit a simple analytic solution, we can still say much about the result using the expedient of the \emph{Wilson flow} \cite{luscher:2010a,luscher:2010b} to infer the existence and uniqueness of a solution. 

\subsection{Interpolation for quantum nonabelian gauge fields}\label{sec:qinterp}
Here we describe how to exploit the classical interpolation procedure derived in the previous subsection to obtain a quantum operation which subdivides, or \emph{zooms into}, a lattice while remaining in the gauge-invariant sector. The resulting quantum interpolation operation is described as a sequence of conditional unitary operations applied to additional ancillary degrees of freedom. For concreteness we focus on the case of two spatial dimensions; the other cases are obvious generalisations. 

The quantum interpolation algorithm proceeds in three steps. The first step is to subdivide each edge of the lattice according to the edge subdivision procedure of Sec.~\ref{sec:paralleltx}:
\begin{center}
	\includegraphics{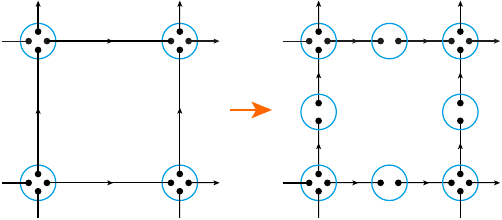}
\end{center} 
If the initial state of the lattice was $|\Psi\rangle$ then the state after the edge subdivision is given by
\begin{equation}
	|\Psi^{(1)}\rangle \equiv \textsl{C}\mathcal{L}^{-1}\left[|\Psi\rangle\otimes \left(\bigotimes_{e\in E}|\omega_0\rangle_{e'}\right)\right],
\end{equation}
where 
\begin{equation}
	\textsl{C}\mathcal{L} \equiv \prod_{e\in E} \textsl{CL}_{e'e}
\end{equation}
and $e'$ denotes the new edge added at the source of the old edge $e$. This isometric procedure doubles the total number of link degrees of freedom and increases the number of vertices by a factor of $3$. We denote the new added vertices at this stage as $V^{(1)}$, i.e., the new vertex set is $V\cup V^{(1)}$.

If the state $|\Psi\rangle$ is the ground state of some hamiltonian $H$ then, after the subdivision step, the state $|\Psi^{(1)}\rangle$ is the ground state of a new hamiltonian $H^{(1)}$ of the form
\begin{equation}\label{eq:parent1}
	H^{(1)} \equiv \textsl{C}\mathcal{L}^\dag \left(H\otimes \mathbb{I} - \sum_{e'} \mathbb{I}\otimes \triangle_{e'}\right) \textsl{C}\mathcal{L},
\end{equation} 
where $-\triangle_{e'}$ has been chosen as an operator acting on the new edge which has $|\omega_0\rangle$ as its unique ground state, although, any other operator which has $|\omega_0\rangle$ as its unique ground state would do. Note that $H^{(1)}$ is now a little more nonlocal than $H$: it contains, via the interpolation unitary, interactions on up to $8$ edges.

The second step is to introduce, at each new lattice point, \emph{two} new gauge-invariant loops in some state $|\psi\rangle$:
\begin{center}
	\includegraphics{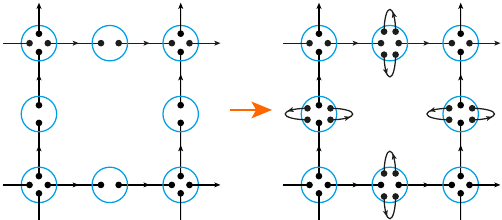}
\end{center} 
The vertex set of the lattice remains unaltered by this step but after this operation the number of link degrees of freedom is increased by $4$ per plaquette (note that we are adding tensor product factors to the \emph{total} hilbert space -- the \emph{physical} hilbert space grows more slowly). The state $|\psi\rangle$ of the new loops is essentially arbitrary. Indeed, in principle, the new loops are allowed to be in some massively entangled state. For illustrative purposes, we consider the state
\begin{equation}
	|\psi\rangle = |U = \mathbb{I}\rangle,
\end{equation} 
with improper wave function $\psi(U) = \delta(U,\mathbb{I})$, which is the optimal choice if we only wish to minimize the curvature per plaquette. The state of the system after this step is
\begin{equation}
	|\Psi^{(2)}\rangle \equiv |\Psi^{(1)}\rangle\otimes \left(\bigotimes_{v'\in V^{(1)}}|\psi\rangle_{v'}|\psi\rangle_{v'}\right).
\end{equation}
We can also describe a hamiltonian which has $|\Psi^{(2)}\rangle$ as its ground state: consider, for example,
\begin{equation}
	H^{(2)} = H^{(1)} + \sum_{e'} (2-\text{Re}(\widehat{u}(e'))),
\end{equation}
where the sum over $e'$ is for all the new edges. This hamiltonian has $|\Psi^{(2)}\rangle$ as its (unique) ground state.

Note that a \emph{normalizable} loop state $|\psi\rangle$ is required to produce a proper isometry. Furthermore, in choosing $|\psi\rangle$, the minimization of curvature is not the only important consideration. For a more realistic example, see (\ref{eq:smeared_loop}).

The final step is to exploit the classical interpolation procedure of the previous subsection to build a conditional unitary operation $CI$ which parallel-transports the ends of each of the loops into the centre of the plaquette. Since this process is a product of operations which plaquettewise commute, we need only describe it for a single plaquette. 
\begin{center}
	\includegraphics{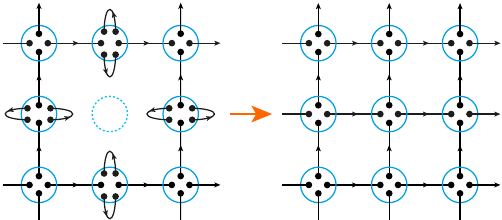}
\end{center} 
We begin the description by denoting the connection variables around the plaquette as $U_0, U_1, \ldots, U_7$:
\begin{center}
	\includegraphics{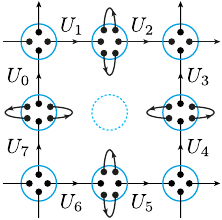}
\end{center} 
Writing $U(\gamma_0)=U_1U_0$, $U(\gamma_1)=U_3^\dag U_2$, $U(\gamma_2)=U_5^\dag U_4^\dag$, and $U(\gamma_3)=U_7 U_6^\dag$, where $\gamma_0$ is the two-edge path from the vertex located west of the plaquette centre to the vertex north of the plaquette centre and, similarly, $\gamma_1$, $\gamma_2$, and $\gamma_3$ are the paths from north to east, east to south, and south to west, respectively. We then denote by 
\begin{equation}
	A_j, \quad j = 0, 1, 2, 3,
\end{equation}
the elements of $G$ found using the interpolation procedure of the previous subsection applied to the tuple $(U(\gamma_0), U(\gamma_1), U(\gamma_2), U(\gamma_3))$. Finally, we introduce the controlled unitary 
\begin{equation}
	\textsl{CI} \equiv \int |\mathbf{U} \rangle\langle \mathbf{U}|\otimes L_{A_0}\otimes L_{A_1} \otimes R_{A_2} \otimes R_{A_3}  \,d\mathbf{U},
\end{equation}
where $\mathbf{U} \equiv (U_0, U_1, \ldots, U_7)$.  This operation parallel-transports the ends of the new loops into the centre of the plaquette. The state at the end of the third stage is then
\begin{equation}
	|\Psi'\rangle = \textsl{C}\mathcal{I}|\Psi^{(2)}\rangle,
\end{equation}
where
\begin{equation}
	\textsl{C}\mathcal{I}= \prod_{\square} \textsl{CI},
\end{equation}
where the product is taken over the plaquettes (with the newly subdivided edges and additional vertex loops). We denote the vertices added at this stage by $V^{(2)}$. The total vertex set for the new lattice is written $V' = V\cup V^{(1)} \cup V^{(2)}$.

A hamiltonian which has $|\Psi'\rangle$ as its ground state is simply
\begin{equation}\label{eq:parent3}
	H' \equiv \textsl{C}\mathcal{I} H^{(2)} \textsl{C}\mathcal{I}^\dag.
\end{equation}
This hamiltonian has, in general, interactions involving up to $12$ edges.

The entire quantum interpolation procedure may be written as a gigantic isometry
\begin{equation}
	|\Psi'\rangle \equiv \mathcal{V}_{\psi} |\Psi\rangle,
\end{equation}
for a proper choice of the wavefunction $|\psi\rangle$ for the added vertex loops.

\subsection{The renormalisation group}
Any isometry of the form $\mathcal{V}:\mathcal{H}_A\rightarrow \mathcal{H}_A\otimes \mathcal{H}_E$ may be written as the action of some unitary operation $\mathcal{U}$ acting on some ancillary state $|0\rangle$ of $\mathcal{H}_E$: 
\begin{equation}
	\mathcal{V} |\psi\rangle_A = \mathcal{U}|\psi\rangle_A|0\rangle_E, \quad \forall |\psi\rangle_A.
\end{equation}
Corresponding to any such isometry there is a \emph{completely positive map} (CP map) or \emph{channel} $\mathcal{E}$ given by
\begin{equation}
	\mathcal{E}(\rho_{AE}) \equiv \mathcal{I}_{A} \otimes \tr_E ( \mathcal{U}^\dag \rho_{AE} \mathcal{U} )
\end{equation}
which admits the interpretation as a \emph{coarse graining} operation or \emph{renormalisation} \footnote{Note that $\mathcal{E}$ is \emph{not} the usual CP map $\mathcal{F}$ arising from the isometry $\mathcal{V}$, rather, it is the \emph{inverse} or \emph{transpose channel} of the CP map $\mathcal{F}$.}.  This channel exactly \emph{undoes} the interpolation $\mathcal{V}$: 
\begin{equation}
	\mathcal{E}(\mathcal{V}|\psi\rangle_A\langle\psi| \mathcal{V}^\dag) = |\psi\rangle_A\langle\psi|, \quad \forall |\psi\rangle_A.
\end{equation}
In our present situation the corresponding completely positive map $\mathcal{E}$ may be interpreted as a Migdal-Kadanoff block renormalisation transformation (note that our transformation crucially differs from the original Migdal-Kadanoff RG \cite{kadanoff:1976a,migdal:1975b,migdal:1975a} by including a \emph{disentanglement} step \cite{vidal:2006a,vidal:2007a} --- the interpolation operation). To produce the channel we simply collect the unitary operations we applied throughout the quantum interpolation procedure and apply them in reverse while tracing out the ancillary degrees of freedom introduced during the interpolation:
\begin{equation}
	\mathcal{E}(\cdot) \equiv   \tr_{E^{(1)}}\left[\textsl{C}\mathcal{L}\left(\textsl{C}\mathcal{I}^\dag(\cdot)\textsl{C}\mathcal{I}  \right )\textsl{C}\mathcal{L}^\dag\right],
\end{equation}
where $E^{(1)}$ denotes the added edges (after having been parallel-transported to their source vertices).

One verifies that both lattice fixed points are fixed by this RG coarse-graining transformation, i.e.,
\begin{equation}
	\mathcal{E}(|\Omega(\infty)\rangle\langle\Omega(\infty)|) = |\Omega(\infty)\rangle\langle\Omega(\infty)|,
\end{equation}
and
\begin{equation}
	\mathcal{E}(|\Omega(0)\rangle\langle\Omega(0)|) = |\Omega(0)\rangle\langle\Omega(0)|.
\end{equation}
Hence the quantum interpolation procedure is an \emph{inverse} to our Migdal-Kadanoff RG $\mathcal{E}$.

Note that, although the map $\mathcal{E}$ manifestly preserves the gauge invariance of any gauge-invariant input state, in general the result is a gauge-invariant \emph{mixed} state which, furthermore, need not be decomposable as a classical mixture of gauge-invariant pure states \footnote{This is a consequence of the coarse-grained state $\rho$ not generally being invariant under ``non-diagonal'' gauge transformations $\rho \neq U_{\{x\}}\rho U_{\{x'\}}^\dagger$, where $U_{\{x\}}$ is the unitary that implements the gauge transformation parameterised by a set of gauge group elements $\{x\}$.}, a result of tracing out a part of the lattice where there is nontrivial curvature \cite{livine_2014}. Indeed, although $\mathcal{E}$ is an appropriate coarse-graining map for states $\mathcal{V}|\psi\rangle_A$ produced by our fine-graining ansatz, we have no reason to expect it to be a good coarse-graining map for all input states.

One can reinterpret the RG maps described here in the context of \emph{multigrid methods} \cite{wesseling:2004a}. Here the problem is that the solution of a discretisation of a continuous equation of motion usually takes a number of computational steps scaling as the inverse of the lattice spacing. This slowdown can be understood physically: numerical solvers such as Gauss-Seidel work by eliminating high momentum degrees of freedom, i.e., they ``refrigerate" the system. As the lattice spacing is decreased the corresponding energy of infrared degrees of freedom effectively decreases (we are putting more degrees of freedom in the same interval in momentum space). Multigrid methods exploit a coarse lattice to first solve for the infrared degrees of freedom and then they interpolate the solution onto a finer lattice via a \emph{prolongation} or \emph{interpolation} map where the solution is then refined to correct the UV degrees of freedom. In this sense one can directly interpret our quantum interpolation scheme as a quantum prolongation map and the RG map $\mathcal{E}$ as an \emph{averaging map}.

\section{The ground-state ansatz}
\label{sec:gs-ansatz}

\subsection{The basic ansatz}
The quantum interpolation procedure described in the previous section may be exploited to write down a contractible tensor network state, a MERA, which satisfies properties 1., 2., 3., and, to some extent, 5.\ of Sec.~\ref{sec:problem}. The basic idea is simple: recursively define
\begin{equation}\label{eq:basicansatz}
	|\Psi_m\rangle = \textsl{C}\mathcal{V}|\Psi_{m-1}\rangle, \quad m = 1, 2, \ldots,
\end{equation}
with $|\Psi_0\rangle \equiv |\Omega(\infty)\rangle$. Because $|\Psi_m\rangle$ may be written as a sequence of a local introduction of ancillary product states followed by local unitaries it acquires the structure of a multi-scale entanglement renormalisation ansatz (MERA). This immediately entails a \emph{contractibility} guarantee: the computation of all $n$-pt correlation functions may be efficiently --- and analytically --- carried out. 

Properties 1., 2., and 3., of Sec.~\ref{sec:problem} are true by construction. However, we will have to work harder to establish properties 4.\ and 5. There are a couple of undesirable features of the family $|\Psi_m\rangle$: (a) the ansatz isn't manifestly invariant under euclidean symmetries; (b) all Wilson loops have a zero expectation value; and (c) the family is an \emph{ansatz}, i.e., it is not \emph{exactly} the ground state of $H(g_H)$ for any value of $g_H$. These defects can be corrected in at least three different ways via: (i) producing a more complicated tensor network by exploiting a tool from the analysis of strongly interacting quantum spin systems, namely \emph{quasi-adiabatic continuation}; (ii) exploiting a more complicated initial state $|\Psi_0\rangle$; or (iii) carefully adjusting the ancillary states $|\psi\rangle$ introduced at each stage. It turns out that these three expedients are essentially equivalent, however, they have different strengths and weaknesses. We'll explore them in turn in the following subsections.  

\subsection{Improving the ansatz: quasi-adiabatic continuation}
In the first approach to improve our ansatz we just directly construct the exact ground state of $H_{\text{KS}}(g_H)$. This is achieved by employing a procedure known as \emph{quasi-adiabatic continuation} \cite{hastings:2005a, osborne:2006a}. The way this works is as follows. Suppose we've exactly constructed the ground state $|\Omega(g)\rangle$ for some non-infinite value of $g$. We now subdivide this state according to the quantum interpolation procedure to produce $|\Psi'\rangle = \textsl{C}\mathcal{V}_{\psi}|\Omega(g)\rangle$. We \emph{conjecture} that $|\Omega(g')\rangle$ is an approximation to the ground state of $H_{\text{KS}}(g'_H)$ for $g'_H < g_H$: this is entirely plausible as the interpolation algorithm exactly minimises the potential energy of the new degrees of freedom at the expense of the kinetic energy so the resulting state should be closer to the ground state of $H_{\text{KS}}(g_H)$ with a smaller value of $g_H$. The state $|\Psi'\rangle$ has a \emph{parent hamiltonian} $H'$ given by the prescription described in Sec.~\ref{sec:qinterp}. We now write 
\begin{equation}
	H(g'_H) = H' + \Delta',
\end{equation}
and regard $\Delta'$ as a \emph{perturbation}. We now make a crucial second \emph{conjecture}, namely that $H'$ is in \emph{the same phase} as $H(g'_H)$. A limited form of this conjecture follows from a generalisation of the methods of \cite{bravyi:2010b, bravyi:2010c, michalakis:2013a}; one needs to develop a generalised Lieb-Robinson bound \cite{lieb:1972a, hastings:2005b, nachtergaele:2010b} on the propagation of quantum correlations which applies to our specific setting: while the potential energy term is a sum of bounded operators the kinetic energy term is a sum of unbounded operators, and the standard Lieb-Robinson bound doesn't apply. (Such a bound may be found by generalising the construction of \cite{premont-schwarz:2010a}.) Establishing our second conjecture in full generality would require techniques going beyond those developed for the study of the stability of quantum phases. One key feature which acts in our favour here is that we are free to modify the Kogut-Susskind hamiltonian by the addition of arbitrary UV irrelevant terms: this relaxes our task by allowing us to explore more general adiabatic paths.

If our second conjecture is correct then we can adiabatically deform $H'$ to $H(g'_H)$ without closing the gap and encountering a quantum phase transition. Thus we construct the interpolating hamiltonian 
\begin{equation}
	K(s) = H' + (1-s)\Delta'.
\end{equation}
The next step is to exploit the adiabatic path $K(s)$, $s\in [0, 1]$ to construct the unitary quasi-adiabatic continuation process $\mathcal{U}$. This unitary is approximately local (this follows from a generalisation of the arguments presented in \cite{osborne:2005d}) and has the property that $\mathcal{U}|\Psi'\rangle = |\Omega(g'_H)\rangle$.  Remarkably, thanks to the locality of $\mathcal{U}$, the combination $\mathcal{U}\textsl{C}\mathcal{V}_{\psi}$ retains the structure of (a layer of) a MERA tensor network.

Our improved ansatz is given by
\begin{equation}\label{eq:imprvansatz}
	|\Psi_m\rangle = \mathcal{U}_{m-1}\textsl{C}\mathcal{V}|\Psi_{m-1}\rangle, \quad m = 1, 2, \ldots,
\end{equation}
and acquires, thanks to the quasi-locality of $\mathcal{U}_{m-1}$, the structure of a MERA tensor network.

The procedure we've just described yields a contractible tensor network: we simply iterate the process to obtain a representation for $|\Omega(g)\rangle$ for a decreasing sequence $\{g_m\}_{m=1}^\infty$ of values. Does the sequence $\{g_m\}_{m=1}^\infty$ converge to zero? This is rather far from obvious --- it is entirely plausible that $\lim_{m\rightarrow \infty} g_m = g^\star >0$. However, such a situation would  contradict asymptotic freedom, which is understood to apply in the case under consideration here. Thus, although it is a conjecture that $\lim_{m\rightarrow \infty} g_m = 0$, it is one that already has considerable evidence. To prove this conjecture one would need to exclude the possibility of a value of $g$ such that $\textsl{C}\mathcal{V}_{\psi}|\Omega(g)\rangle = |\Omega(g)\rangle$. This is an eigenvector equation and a simple energy argument should likely settle the issue. 

The price we pay for the improved results of the corrected ansatz is that the causal cone is widened (indeed, it becomes infinite). This is not immediately a problem as the cone is effectively finite, with exponentially damped tails. However it does become more computationally demanding to extract expectation values.

The quasi-adiabatic correction procedure may be regarded as compensating for the errors made in ignoring the kinetic energy term of the Kogut-Susskind hamiltonian when the interpolation is carried out. Indeed, without correction, it is easy to see that the expectation value of the kinetic energy operator on the new edges diverges. One might imagine that there is a simpler way to modify the ansatz by exploiting the \emph{heat flow} generated by the kinetic energy, however, unfortunately, because the heat flow is not unitary, this will not directly lead to a contractible tensor network as the causal structure is lost. 

\subsection{Improving the ansatz: more complicated initial states}
Here we explore a second approach to improving our basic ansatz. The physical motivation here comes from regarding our quantum interpolation procedure as an \emph{ultraviolet completion} whereby the state of an initial coarse lattice with correposponding momentum cutoff $\Lambda$ is refined, shifting the UV cutoff upwards. According to this interpretation we require that the initial state faithfully encodes the large-scale low-energy physics of Yang-Mills theory; the UV completion provided by the quantum interpolation procedure then takes care of the high-energy small-scale processes. Here we are helped out by the observation that at large scales quantum Yang-Mills theory is very close to the strong-coupling fixed point. 

Our basic ansatz is simply the case where the initial state is at \emph{lattice} strong coupling. However, this is not sufficient to give a good enough approximation because there are no correlations and when the UV completion is carried out large-scale Wilson loops are identically zero. This can be corrected by slightly moving away from lattice strong coupling; we consider instead the initial state
\begin{equation}
	|\Psi_0\rangle \equiv \lim_{\beta\rightarrow\infty} \frac{e^{-\beta H_{\text{KS}}(\epsilon)}|\Omega_\infty\rangle}{\|e^{-\beta H_{\text{KS}}(\epsilon)}|\Omega_\infty\rangle\|^{\frac12}},
\end{equation}    
where $\epsilon > 0$ is small but nonzero. Now it follows from a generalisation of the arguments of \cite{bravyi:2010c,michalakis:2013a} that there is a nonzero $\epsilon$ such that $H_{\text{KS}}(\epsilon)$ is \emph{adiabatically connected} to $H_{\text{KS}}(0)$. By again employing a quasi-adiabatic continuation process we can infer that $|\Psi_0\rangle$ is a contractible tensor network: this follows from a straightforward generalisation of the arguments of \cite{osborne:2006a,hastings:2005a}. 

After the improved initial state is obtained we simply apply the quantum interpolation isometry to send the lattice spacing to zero. We are now guaranteed that infrared contributions to the correlation functions are well represented. Indeed, it follows that wilson loops now enjoy an area law scaling. 

\subsection{Improving the ansatz: smoother interpolation}
A final approach to improving the basic ansatz is to make smaller interpolation steps. The problem is that our interpolation procedure is somewhat discontinuous because it effects a scale change by a discrete factor of $2$ and this introduces UV artifacts that need to be corrected by quasi-adiabatic continuation. A possible way to deal with this is to make smaller scale changes. Indeed, the optimal situation would be to make an \emph{infinitesimal} scale change. Making smaller scale changes could be carried out by studying the curvature interpolation problem where instead of subdiving a single plaquette we take $m$ plaquettes and replace them, in as smooth a way as possible, by $n>m$ plaquettes. We haven't had much success in formalising this intuition yet, and leave it as an interesting open problem for the reader.

\subsection{Expectation values}
Here we describe how to compute the expectation values of local operators and string operators for both the basic ansatz Eq.~(\ref{eq:basicansatz}) and the improved ansatz Eq.~(\ref{eq:imprvansatz}). The discussion in this section is framed in the setting of a lattice in two spatial dimensions. The results we present here are, however, representative of the somewhat more involved general case.

There are a variety of observables relevant for large-scale low-energy physics. We mostly focus on gauge-invariant observables arising as the discretisations of quantum field operators. The first class of operator is the lattice magnetic field operator given by a Wilson loop around an elementary plaquette in the refined lattice:
\begin{equation}
	\tr(\widehat{u}_{\cwplaqsub}),
\end{equation}
where $\cwplaqsub$ denotes a specific plaquette.  The way we compute the expectation value of such an operator in the state Eq.~(\ref{eq:basicansatz}) is to compute the transformation of $\tr(\widehat{u}_{\cwplaqsub})$ under the action of $ \textsl{C}\mathcal{V}$. It turns out that, since $\tr(\widehat{u}_{\cwplaqsub})$ is \emph{diagonal} in the position basis, this transformation may be deduced somewhat indirectly from the classical interpolation procedure, the result is:
\begin{equation}
	\textsl{C}\mathcal{V}^\dag\tr(\widehat{u}_{\cwplaqsub})\textsl{C}\mathcal{V} =  \lambda^2 \tr(\widehat{u}_{\cwplaqsub'}^{\frac14}),
\end{equation}
where $\lambda = \frac12 \langle \psi|\tr(\widehat{u})|\psi\rangle$ and $|\psi\rangle$ is the auxiliary state introduced during the quantum interpolation procedure and $\cwplaqsub'$ represents the image of the plaquette $\cwplaqsub$ in the coarse-grained lattice. The coefficient $\lambda$ ranges from $0$, for $|\psi\rangle \equiv |\omega_0\rangle$, to $1$ in the case $|\psi\rangle \equiv |U \equiv \mathbb{I}\rangle$. When $\lambda \not=1$ the action of interpolation becomes somewhat intricate and we'll present the details of this case in a separate paper. The case where $\lambda = 1$, which represents an interpolation so that the interpolated variables are as flat as possible, is somewhat simpler. Here we can readily iterate the procedure: the \emph{ascending channel}, defined by
\begin{equation}
	\mathcal{A}(M_\square) \equiv \textsl{C}\mathcal{V}^\dag M_{{\square}'} \textsl{C}\mathcal{V},
\end{equation}
where $\square'$ is the image of $\square$ in the coarse-grained lattice,
acts in a particularly simple way on arbitrary (smooth) functions of plaquette operators:
\begin{equation}
	\mathcal{A}(f(\tr(\widehat{u}_{\cwplaqsub})) = f(\tr(\widehat{u}^{\frac14}_{\cwplaqsub'})).
\end{equation}
This allows us to determine that the flux operator 
\begin{equation}
	\widehat{\Phi}_\square \equiv \arccos\left(\frac12\tr(\widehat{u}_{\cwplaqsub} )\right)
\end{equation}
is an eigenoperator of the CP map $\mathcal{A}$, specifically,
\begin{equation}
	\mathcal{A}(\widehat{\Phi}_\square) = \frac14 \widehat{\Phi}_{\square'}.
\end{equation}
We also find that products transform in a simple way, e.g.,
\begin{equation}\label{eq:multiinterp}
	\mathcal{A}(\widehat{\Phi}_{\square_1} \cdots \widehat{\Phi}_{\square_l}) = \frac{1}{4^l} \widehat{\Phi}_{\square_1'} \cdots \widehat{\Phi}_{\square_l'}.
\end{equation}
where $\square_j$, $j=1, \ldots, l$, is a collection of $l$ plaquettes. In general, for $\lambda \not= 1$, this is not the case and the action of $\mathcal{A}$ is substantially more complicated. 

The expectation value of, e.g., $\widehat{\Phi}_\square$ in the state $|\Psi_m\rangle$ is now readily computed by recursion, we find that
\begin{equation}
	\begin{split}
	\langle\Psi_m|\widehat{\Phi}_\square|\Psi_m\rangle &= \frac{1}{4^m} \langle \omega_0|\arccos\left(\frac12\tr(\widehat{u})\right)|\omega_0\rangle \\
	&= \frac{1}{4^m}\frac{\pi}{2}.
	\end{split}
\end{equation}

We can also readily calculate the $n$-point correlation functions of observables such as $\widehat{\Phi}_\square$ in the $\lambda = 1$ case. For example, consider two such operators: using Eq.~(\ref{eq:multiinterp}), we readily deduce that their $2$-point function is given by 
\begin{equation}
	\langle \Psi_m|\widehat{\Phi}_\square\widehat{\Phi}_{\square'}|\Psi_m \rangle = \begin{cases} \frac{1}{4^{2m}}\left(\frac{\pi^2}{3} - \frac12\right), &\quad \square \sim\square' \\ \frac{1}{4^{2m}}\frac{\pi^2}{4}, &\quad \text{otherwise}, \end{cases}
\end{equation}
where $\square \sim \square'$ means that both plaquettes are situated within the same coarse plaquette.

\section{Numerical examples}

\begin{figure}[b]
  \includegraphics{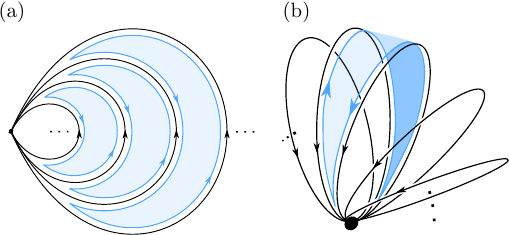}
  \caption{\label{fig:earring} Two possible embeddings (a) and (b) of the graph associated with the
           Hamiltonian lattice principal chiral field model.
           For the gauge groups $G=U(1)$ and $G=SU(2)$ this model is equivalent to the
           (1+1)-dimensional quantum rotor model
           for the rotation groups $O(2)$ and $O(4)$, respectively.
           }
\end{figure}

\begin{figure*}[ht]
  \includegraphics[width=0.48\linewidth]{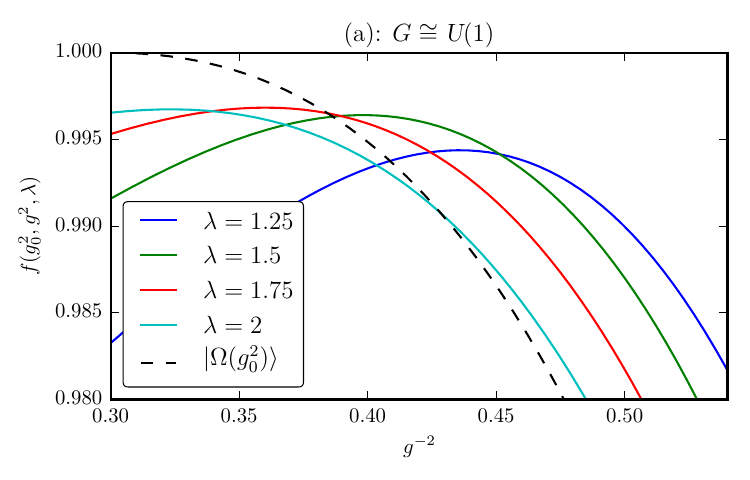}
  \includegraphics[width=0.48\linewidth]{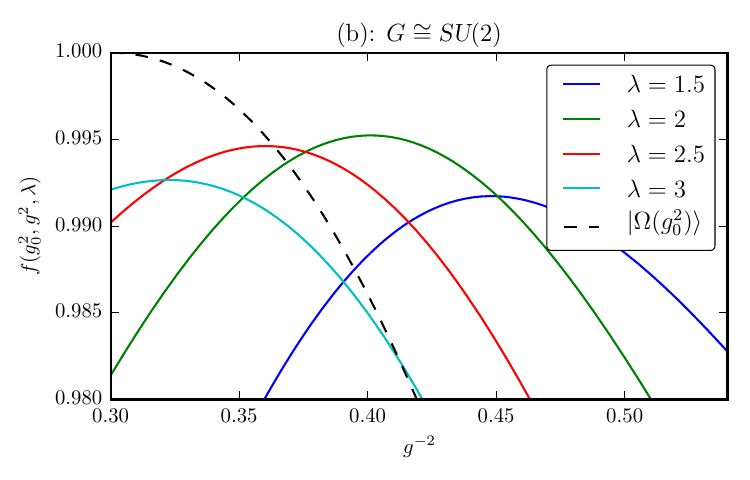}
  \caption{\label{fig:interp-fid} The fidelity per site $f$ of the 
    MPS ground state of the $G \cong \uone$ and $G \cong \su2$ 
    principal chiral field models (\ref{eq:H_R}) at inverse coupling $g^{-2}$ 
    with the fine-grained state $V(\lambda) |\Omega(g_0^{-2})\rangle$ for
    $g_0^{-2} = 0.3$. The initial state of the added loops is $|\psi(\lambda)\rangle$. 
    As a reference, we also plot the fidelity with $|\Omega(g_0^{-2})\rangle$.
    MPS ground states have bond dimension and Fourier mode cutoff $D=22$ and $|n| \le 5$
    for $G \cong \uone$, and $D=14$ and $l \le \frac{3}{2}$ for $G \cong \su2$. The
    interpolated states have $D=64$ and $D=30$, respectively.
    We see that $f(g_0^{-2}, g^{-2}, \lambda)$ has a clear
    maximum for $g^{-2} > g_0^{-2}$, showing
    that the fine-grained state is closer to the ground state at a larger
    inverse coupling than to the original ground state at $g_0^{-2}$. Furthermore, 
    the fine-grained states are often significantly better approximations than 
    $|\Omega(g_0^{-2})\rangle$ for smaller values of $\lambda$.
    Note that some of the error in the fine-grained states is due to 
    finite-entanglement effects and the Fourier-mode cutoff, particularly
    at larger~$g^{-2}$ and for $G \cong \su2$.
}
\end{figure*}

Here we present some preliminary results for interpolation of the $(1+1)$D principal chiral field model with $G \cong \uone$ and $G \cong \su2$ on a spatial lattice (a chain of $N$ sites). This model has the Kogut-Susskind Hamiltonian
\begin{equation} \label{eq:H_R}
	H_R(g) = -\frac{g^2}{2 a} \sum_e \Delta_e + \frac{1}{g^2 a}\left( 2 - \sum_e \text{Re}(\tr (\widehat{u}_e \widehat{u}_{e+1}^\dagger)) \right),
\end{equation}
and is, insofar as we do not restrict to gauge-invariant states, equivalent to the $O(2)$ ($O(4)$) quantum rotor models in $(1+1)$D~\cite{sachdev:2011a} in case $G \cong \uone$ ($G \cong \su2$). The graph describing the gauge-symmetry of~(\ref{eq:H_R}) is visualized in Figure~\ref{fig:earring}.  As noted in Section~\ref{sec:gis}, on such a graph the gauge symmetry is reduced to a global $G$~symmetry, or no symmetry at all in case of abelian~$G$. $H_R$~is further invariant under global rotations $(L_g)^{\otimes N}$ and $(R_g)^{\otimes N}$ $\forall g \in G$. Note that, unlike local gauge symmetry, these global symmetries can be spontaneously broken, albeit not in the ground state (a consequence of the Coleman-Mermin-Wagner theorem, see for example~\cite{peskin:1995a}).

The $G \cong \uone$ model is known to have a confining phase at large $g$ in which the Hamiltonian has a spectral gap between the ground state and the first excited state, as well as a deconfined, gapless phase at small $g$ in which correlations decay algebraically, and which is described by a free bosonic conformal field theory (central charge $c=1$). The transition separating the two phases is of the Kosterlitz-Thouless type~\cite{sachdev:2011a}. The nonabelian case of $G \cong \su2$ behaves very differently, having only one confining phase that persists across strong and weak couplings~\cite{sachdev:2011a}. However, despite the lack of a phase transition there is a \emph{crossover} region separating weak and strong coupling in which the scaling with $g$ of quantities such as the mass gap changes rapidly. There is much evidence~\cite{creutz:1985a} suggesting that this phase structure, including the crossover phenomenon, is shared by $(3+1)$D lattice Yang-Mills theory for nonabelian~$G$. Similarly, the $G \cong \uone$ phase diagram is similar to that of $(3+1)$D $\uone$ lattice gauge theory (pure, compact QED), except that the phase transition of the latter theory is thought to be first order~\cite{arnold:2003a}. Despite the lack of local gauge symmetry, these similarities to Yang-Mills theory make $H_R$ a useful toy model for testing the fine-graining ansatz presented in this paper.

Using the numerical framework developed in \cite{milsted_matrix_2016}, in which $H_R$ was studied using infinite matrix product states (MPS) as a variational class of states, we test curvature interpolation by comparing MPS ground states before and after fine-graining. 
To represent states in $\mathcal{H} = (L^2(G))^{\otimes N}$ as MPS with a finite-dimensional effective local Hilbert space, we use the Fourier modes of $G$ as a basis for $L^2(G)$ and implement a momentum cutoff specified by the label $l_{\max}$ of the irrep with the maximum allowed kinetic energy. Note that, although this cutoff is needed for the present numerical checks, it is not a requirement of the analytical framework outlined in the preceding sections.

The fine-graining isometries 
\begin{equation} \label{eq:interp_R}
  \mathcal{V}_\psi = \prod_{e} \textsl{C}I_{e,e',e + 1} |\psi\rangle_{e'},
\end{equation}
consist of adding a new uncorrelated loop $e'$ in some state $|\psi\rangle$, situated between the existing loops $e$ and $e+1$ of the chain, 
and then applying an interpolation map 
\begin{align}
  \textsl{C}I_{e,e',e + 1} = \int &|U_e\rangle \langle U_e | \;  \otimes R^\dagger_{A(U_e, U_{e+1})} \\
                        &\otimes \; |U_{e+1}\rangle \langle U_{e+1}| \; dU_e dU_{e+1}, \nonumber
\end{align}
using the classical solution to the interpolation problem 
\begin{equation}
  A(U_e, U_{e+1}) \equiv  \left(U_e \sqrt{U_e^\dagger U_{e+1}} \right),
\end{equation}
which is the SLERP result (\ref{eq:slerp}) for $n=2$.
A candidate $|\psi\rangle$ that is compatible with the Fourier cutoff is the gauge-invariant single-loop state
\begin{equation} \label{eq:smeared_loop}
  |\psi(\lambda)\rangle \propto \sum_{l}^{\text{cutoff}} e^{-\lambda |l|}\; |\mathbb{I}\rangle_l,
\end{equation}
where $l = [-l_{\max}, l_{\max}]$ for $\uone$ and $l = 0, \frac{1}{2},\dots, l_{\max}$ for $\su2$. $|\mathbb{I}\rangle_l$ is the irrep-$l$ representation of the identity element of $G$. To further mitigate errors due to the cutoff, we additionally set $\langle \psi(\lambda) | \mathbb{I}\rangle_{l_{\max}} = 0$.

To evaluate this fine-graining ansatz we apply $\mathcal{V}(\lambda)~\equiv~\mathcal{V}_{\psi(\lambda)}$ to an MPS ground state $|\Omega(g_0^{-2})\rangle$ of $H_R$ at inverse coupling $g_0^{-2}$ and compute the fidelity per site
\begin{equation}
  f(g_0^{-2}, g^{-2}, \lambda) \equiv \lim_{N \rightarrow \infty} \frac{1}{N} \langle\Omega(g^{-2})|\mathcal{V}(\lambda)|\Omega(g_0^{-2})\rangle,
\end{equation}
of the fine-grained state $\mathcal{V}(\lambda)|\Omega(g_0^{-2})\rangle$ and the MPS ground state $|\Omega(g^{-2})\rangle$ for a range of $g^{-2} > g_0^{-2}$. If curvature interpolation is a good fine-graining ansatz we should find that $f(g_0^{-2}, g^{-2}, \lambda)$ has a peak for some $g^{-2} > g_0^{-2}$ for fixed $g_0^{-2}$, $\lambda$. 

Figure~\ref{fig:interp-fid} shows our results for $G~\cong~\uone$ and $G~\cong~\su2$. In both cases we find values of $\lambda$ for which $\mathcal{V}(\lambda)|\Omega(g_0^{-2})\rangle$ is a significantly better approximation than $|\Omega(g_0^{-2})\rangle$ to MPS ground states at larger inverse couplings, showing promise for curvature interpolation as a fine-graining ansatz. However, errors are clearly introduced, and we observe a corresponding shift in the ground state energy density of the fine-grained states, compared to the closest true ground state. It may be necessary to repair these errors to obtain a truly useful fine-graining ansatz, as described in Section~\ref{sec:gs-ansatz}. It also remains to be seen how well the curvature interpolation functions in the crossover region of nonabelian theories.

\section{The continuum limit}
In this section we sketch how to build a hilbert space associated with the continuum limit of quantum lattice gauge theory. Some of the mathematical details are involved and are postponed to a future paper, however the core idea can be expressed succinctly. 

Conditioned on the validity of the previously mentioned conjectures, we've produced a sequence of states $|\Psi_m\rangle$ which tend, in the limit, to the (lattice) zero-coupling state. The important observation here is that since each term in our sequence is a MERA the correlation length of $|\Psi_m\rangle$ is given by $\xi_m = a_m\lambda^m$, for some $\lambda>0$.  Given that the correlation length $\xi$ of pure Yang-Mills theory is only determined up to a constant which is ultimately fixed by experiment forces us to set the lattice spacing $a_m$ of the state $|\Psi_m\rangle$ to $a_m = a_0\lambda^{-m}$, where $a_0$ is a constant. Thus we have a sequence of states $|\Psi_m\rangle$ for lattices of ever finer discretisation. Each term in the sequence is the result of an isometry applied to a previous term. Further, we can compute all $n$-point functions for this sequence. It turns out that this is enough data to specify a continuum hilbert space and a canonical continuum ground state. 

The continuum hilbert space we describe here is known as a \emph{direct limit} of hilbert spaces; in the context we use it here we call this direct limit the \emph{semicontinuous limit} \footnote{This terminology was suggested to us by Vaughan Jones.} to indicate that it doesn't quite correspond to what we might demand of a full continuous quantum Yang-Mills theory. Note that the direct limit is a basic categorical construction (you can read about it further in, e.g., \cite{lang:2002a}). The application of the direct limit to hilbert spaces has a very long history; one early proposal to use the direct limit to model continuum limits can be found in \cite{bimonte_lattices_1996}, but there are surely prior proposals. It is, for example, a standard technique in quantum gravity \cite{ashtekar_representations_1992, ashtekar_representation_1993, dittrich_discrete_2012}. A recent fascinating attempt to use the direct limit to build continuum limits of lattice theories, in particular, conformal field theories, can be found in \cite{jones_unitary_2014}.

Let $\mathcal{D}$ be the directed set of regular partitions of $\mathbb{R}^d$ induced by integer lattices with lattice spacing $a$, i.e., $a\mathbb{Z}^d$. This set is directed by \emph{refinement}, i.e., a partition $Q$ is a \emph{refinement} of $P$, denoted $P \preceq Q$, if every element of $Q$ is a subset of an element of $P$. (A useful mnemonic to remember the ordering is that $Q$ has ``more'' elements than $P$.) We regard every lattice spacing $a$ as giving rise to a \emph{physically different} lattice.
 
Suppose, further, we associate a hilbert space $\mathcal{H}_P$ with each partition $P\in\mathcal{D}$:
\begin{equation}
	P \mapsto \mathcal{H}_P
\end{equation}
such that for every pair $P\preceq Q$ we have an isometry $T_{Q}^P:\mathcal{H}_P\rightarrow \mathcal{H}_Q$ with the property that for all $P\preceq Q\preceq R$
\begin{equation}
	T_{R}^QT_{Q}^P=T_{R}^P,
\end{equation}
and $T^{P}_P = \mathbb{I}$ for all $P\in\mathcal{D}$. This is, of course, precisely the data specifying a direct system of hilbert spaces. Thus we have enough information to build the \emph{direct limit} hilbert space:
\begin{equation}
	\mathcal{H} \equiv \varinjlim \mathcal{H}_P.
\end{equation}
This space is given by the disjoint union 
\begin{equation}
	\biguplus_{P\in\mathcal{D}}\mathcal{H}_P,
\end{equation}
whose elements are pairs $\langle |\phi\rangle, P\rangle$, $P\in\mathcal{D}$ and $|\phi\rangle\in \mathcal{H}_P$, 
modulo the equivalence relation $\langle |\phi\rangle, P\rangle \sim \langle |\psi\rangle, Q\rangle$ if $|\phi\rangle \in \mathcal{H}_P$ and $|\psi\rangle \in \mathcal{H}_Q$ and there is a partition $R\in\mathcal{D}$ with $P\preceq R$ and $Q\preceq R$ with $T_{R}^P|\phi\rangle = T_{R}^Q|\psi\rangle$. This disjoint union is then completed with respect to the norm $\|\cdot\|$ so that
\begin{equation}
	 \mathcal{H} \equiv \left(\biguplus_{P\in\mathcal{D}}\mathcal{H}_P\bigg/\sim\right)^{\|.\|}.
\end{equation} 
This is the \emph{semicontinuous limit}. You should think of the residents of $\mathcal{H}$ as the \emph{UV completions} of states defined in the IR.  

In our specific case we take for the isometries $T^{P}_Q$ connecting a lattice $P$ and its subdivision refinement $Q$ the combined MERA step $\textsl{C}\mathcal{V}_{\psi}$. In the case where $Q$ is the lattice resulting from several subdivisions we simply take the product $\textsl{C}\mathcal{V}_{\psi}$ of the MERA steps, i.e., $T^{P}_Q = \textsl{C}\mathcal{V}_{\psi}\textsl{C}\mathcal{V}_{\psi}\cdots \textsl{C}\mathcal{V}_{\psi}$. It is easy to check that this leads to a direct system of hilbert spaces. The sequence $|\Psi_m\rangle$ of ground states hence all belong to the \emph{same} single equivalence class $[\langle |\Psi_m\rangle, P_m\rangle]$, i.e., the sequence corresponds to exactly one state $|\Omega\rangle \equiv [\langle |\Psi_m\rangle, P_m\rangle]$ in the semicontinuous limit space $\mathcal{H}$. We conjecture that each improved ansatz corresponds to a different state in $\mathcal{H}$, and that each of these states are ground states of the full continuum quantum Yang-Mills theory with different values of~$g$. 

\section{Conclusions}
In this paper we have sketched a programme to understand the ground state and low-energy excited states of quantum Yang-Mills theory in the temporal gauge as a sequence of tensor network states. We showed how the gauge-invariant sector of hamiltonian lattice gauge theory may be completely parametrised using the expedient of \emph{quantum parallel transport}. Then we introduced the idea of \emph{curvature interpolation} to build an RG which interpolates between the lattice zero coupling and infinite coupling ground states. This leads to a sequence of tensor network states as MERA which we hypothesise to be a model for the ground state of continuous quantum Yang-Mills theory. This hypothesis was validated numerically for the $(1+1)$-dimensional principal chiral field model. We rounded up this paper with a quick overview of how to build the continuous limit rigourously via a construction known as the \emph{semicontinuous limit}. 

There are many things still to be done. Most of the assertions we made are conditioned on major conjectures, which are likely to be hard to resolve. Nonetheless, we feel that there is much to be gained by proceeding with this programme. The most important point we'd like to end on is that the TNS ansatz we've promoted here is \emph{efficiently computable}: all $n$-point correlation functions for local operators can be efficiently computed. This seems important enough by itself to merit further study.

\acknowledgments{This work has greatly benefited from conversations with numerous people. A partial list includes  Karel Van Acoleyen, Jutho Haegeman, Luca Tagliacozzo, Henri Verschelde, Frank Verstraete, and Guifre Vidal, amongst many others. This work was supported by the ERC grants QFTCMPS and SIQS, and by the cluster of excellence EXC201 Quantum Engineering and Space-Time Research. This research was supported in part by Perimeter Institute for Theoretical Physics. Research at Perimeter Institute is supported by the Government of Canada through the Department of Innovation, Science and Economic Development Canada and by the Province of Ontario through the Ministry of Research, Innovation and Science.}


\begin{thebibliography}{114}%
\makeatletter
\providecommand \@ifxundefined [1]{%
 \@ifx{#1\undefined}
}%
\providecommand \@ifnum [1]{%
 \ifnum #1\expandafter \@firstoftwo
 \else \expandafter \@secondoftwo
 \fi
}%
\providecommand \@ifx [1]{%
 \ifx #1\expandafter \@firstoftwo
 \else \expandafter \@secondoftwo
 \fi
}%
\providecommand \natexlab [1]{#1}%
\providecommand \enquote  [1]{``#1''}%
\providecommand \bibnamefont  [1]{#1}%
\providecommand \bibfnamefont [1]{#1}%
\providecommand \citenamefont [1]{#1}%
\providecommand \href@noop [0]{\@secondoftwo}%
\providecommand \href [0]{\begingroup \@sanitize@url \@href}%
\providecommand \@href[1]{\@@startlink{#1}\@@href}%
\providecommand \@@href[1]{\endgroup#1\@@endlink}%
\providecommand \@sanitize@url [0]{\catcode `\\12\catcode `\$12\catcode
  `\&12\catcode `\#12\catcode `\^12\catcode `\_12\catcode `\%12\relax}%
\providecommand \@@startlink[1]{}%
\providecommand \@@endlink[0]{}%
\providecommand \url  [0]{\begingroup\@sanitize@url \@url }%
\providecommand \@url [1]{\endgroup\@href {#1}{\urlprefix }}%
\providecommand \urlprefix  [0]{URL }%
\providecommand \Eprint [0]{\href }%
\providecommand \doibase [0]{http://dx.doi.org/}%
\providecommand \selectlanguage [0]{\@gobble}%
\providecommand \bibinfo  [0]{\@secondoftwo}%
\providecommand \bibfield  [0]{\@secondoftwo}%
\providecommand \translation [1]{[#1]}%
\providecommand \BibitemOpen [0]{}%
\providecommand \bibitemStop [0]{}%
\providecommand \bibitemNoStop [0]{.\EOS\space}%
\providecommand \EOS [0]{\spacefactor3000\relax}%
\providecommand \BibitemShut  [1]{\csname bibitem#1\endcsname}%
\let\auto@bib@innerbib\@empty
\bibitem [{\citenamefont {Wilson}(1974)}]{wilson:1974b}%
  \BibitemOpen
  \bibfield  {author} {\bibinfo {author} {\bibfnamefont {K.~G.}\ \bibnamefont
  {Wilson}},\ }\href {\doibase 10.1103/PhysRevD.10.2445} {\bibfield  {journal}
  {\bibinfo  {journal} {Phys. Rev. D}\ }\textbf {\bibinfo {volume} {10}},\
  \bibinfo {pages} {2445} (\bibinfo {year} {1974})}\BibitemShut {NoStop}%
\bibitem [{\citenamefont {Creutz}(1985)}]{creutz:1985a}%
  \BibitemOpen
  \bibfield  {author} {\bibinfo {author} {\bibfnamefont {M.}~\bibnamefont
  {Creutz}},\ }\href@noop {} {\emph {\bibinfo {title} {Quarks, gluons and
  lattices}}}\ (\bibinfo  {publisher} {Cambridge University Press},\ \bibinfo
  {address} {Cambridge},\ \bibinfo {year} {1985})\BibitemShut {NoStop}%
\bibitem [{\citenamefont {D\"urr}\ \emph {et~al.}(2008)\citenamefont {D\"urr},
  \citenamefont {Fodor}, \citenamefont {Frison}, \citenamefont {Hoelbling},
  \citenamefont {Hoffmann}, \citenamefont {Katz}, \citenamefont {Krieg},
  \citenamefont {Kurth}, \citenamefont {Lellouch}, \citenamefont {Lippert},
  \citenamefont {Szabo},\ and\ \citenamefont {Vulvert}}]{duerr:2008a}%
  \BibitemOpen
  \bibfield  {author} {\bibinfo {author} {\bibfnamefont {S.}~\bibnamefont
  {D\"urr}}, \bibinfo {author} {\bibfnamefont {Z.}~\bibnamefont {Fodor}},
  \bibinfo {author} {\bibfnamefont {J.}~\bibnamefont {Frison}}, \bibinfo
  {author} {\bibfnamefont {C.}~\bibnamefont {Hoelbling}}, \bibinfo {author}
  {\bibfnamefont {R.}~\bibnamefont {Hoffmann}}, \bibinfo {author}
  {\bibfnamefont {S.~D.}\ \bibnamefont {Katz}}, \bibinfo {author}
  {\bibfnamefont {S.}~\bibnamefont {Krieg}}, \bibinfo {author} {\bibfnamefont
  {T.}~\bibnamefont {Kurth}}, \bibinfo {author} {\bibfnamefont
  {L.}~\bibnamefont {Lellouch}}, \bibinfo {author} {\bibfnamefont
  {T.}~\bibnamefont {Lippert}}, \bibinfo {author} {\bibfnamefont {K.~K.}\
  \bibnamefont {Szabo}}, \ and\ \bibinfo {author} {\bibfnamefont
  {G.}~\bibnamefont {Vulvert}},\ }\href {\doibase 10.1126/science.1163233}
  {\bibfield  {journal} {\bibinfo  {journal} {Science}\ }\textbf {\bibinfo
  {volume} {322}},\ \bibinfo {pages} {1224} (\bibinfo {year} {2008})},\ \Eprint
  {http://arxiv.org/abs/0906.3599} {arXiv:0906.3599} \BibitemShut {NoStop}%
\bibitem [{Note1()}]{Note1}%
  \BibitemOpen
  \bibinfo {note} {\protect \url
  {http://quantumfrontiers.com/2012/12/11/fundamental-physics-prize-prediction-polyakov/}.}\BibitemShut
  {Stop}%
\bibitem [{\citenamefont {'t~Hooft}(2005)}]{thooft:2005}%
  \BibitemOpen
  \bibinfo {editor} {\bibfnamefont {G.}~\bibnamefont {'t~Hooft}},\ ed.,\
  \href@noop {} {\emph {\bibinfo {title} {50 years of {Y}ang-{M}ills theory}}}\
  (\bibinfo  {publisher} {World Scientific Publishing Co. Pte. Ltd.},\ \bibinfo
  {address} {Singapore},\ \bibinfo {year} {2005})\BibitemShut {NoStop}%
\bibitem [{\citenamefont {Greensite}(1979)}]{greensite:1979a}%
  \BibitemOpen
  \bibfield  {author} {\bibinfo {author} {\bibfnamefont {J.~P.}\ \bibnamefont
  {Greensite}},\ }\href {\doibase 10.1016/0550-3213(79)90178-0} {\bibfield
  {journal} {\bibinfo  {journal} {Nucl. Phys. B}\ }\textbf {\bibinfo {volume}
  {158}},\ \bibinfo {pages} {469 } (\bibinfo {year} {1979})}\BibitemShut
  {NoStop}%
\bibitem [{\citenamefont {Feynman}(1981)}]{feynman:1981a}%
  \BibitemOpen
  \bibfield  {author} {\bibinfo {author} {\bibfnamefont {R.~P.}\ \bibnamefont
  {Feynman}},\ }\href {\doibase http://dx.doi.org/10.1016/0550-3213(81)90005-5}
  {\bibfield  {journal} {\bibinfo  {journal} {Nucl. Phys. B}\ }\textbf
  {\bibinfo {volume} {188}},\ \bibinfo {pages} {479 } (\bibinfo {year}
  {1981})}\BibitemShut {NoStop}%
\bibitem [{\citenamefont {Karabali}\ \emph {et~al.}(1998)\citenamefont
  {Karabali}, \citenamefont {Kim},\ and\ \citenamefont
  {Nair}}]{karabali:1998a}%
  \BibitemOpen
  \bibfield  {author} {\bibinfo {author} {\bibfnamefont {D.}~\bibnamefont
  {Karabali}}, \bibinfo {author} {\bibfnamefont {C.}~\bibnamefont {Kim}}, \
  and\ \bibinfo {author} {\bibfnamefont {V.~P.}\ \bibnamefont {Nair}},\ }\href
  {\doibase http://dx.doi.org/10.1016/S0370-2693(98)00751-5} {\bibfield
  {journal} {\bibinfo  {journal} {Phys. Lett. B}\ }\textbf {\bibinfo {volume}
  {434}},\ \bibinfo {pages} {103 } (\bibinfo {year} {1998})}\BibitemShut
  {NoStop}%
\bibitem [{\citenamefont {Samuel}(1997)}]{samuel:19967a}%
  \BibitemOpen
  \bibfield  {author} {\bibinfo {author} {\bibfnamefont {S.}~\bibnamefont
  {Samuel}},\ }\href {\doibase 10.1103/PhysRevD.55.4189} {\bibfield  {journal}
  {\bibinfo  {journal} {Phys. Rev. D}\ }\textbf {\bibinfo {volume} {55}},\
  \bibinfo {pages} {4189} (\bibinfo {year} {1997})}\BibitemShut {NoStop}%
\bibitem [{\citenamefont {Auerbach}(1994)}]{auerbach:1994a}%
  \BibitemOpen
  \bibfield  {author} {\bibinfo {author} {\bibfnamefont {A.}~\bibnamefont
  {Auerbach}},\ }\href@noop {} {\emph {\bibinfo {title} {Interacting electrons
  and quantum magnetism}}}\ (\bibinfo  {publisher} {Springer-Verlag},\ \bibinfo
  {address} {New York},\ \bibinfo {year} {1994})\BibitemShut {NoStop}%
\bibitem [{\citenamefont {Sachdev}(2011)}]{sachdev:2011a}%
  \BibitemOpen
  \bibfield  {author} {\bibinfo {author} {\bibfnamefont {S.}~\bibnamefont
  {Sachdev}},\ }\href@noop {} {\emph {\bibinfo {title} {Quantum phase
  transitions}}},\ \bibinfo {edition} {2nd}\ ed.\ (\bibinfo  {publisher}
  {Cambridge University Press},\ \bibinfo {address} {Cambridge},\ \bibinfo
  {year} {2011})\BibitemShut {NoStop}%
\bibitem [{\citenamefont {Schollw{\"o}ck}(2005)}]{schollwoeck:2005a}%
  \BibitemOpen
  \bibfield  {author} {\bibinfo {author} {\bibfnamefont {U.}~\bibnamefont
  {Schollw{\"o}ck}},\ }\href {\doibase 10.1103/RevModPhys.77.259} {\bibfield
  {journal} {\bibinfo  {journal} {Rev. Modern Phys.}\ }\textbf {\bibinfo
  {volume} {77}},\ \bibinfo {pages} {259} (\bibinfo {year} {2005})},\ \Eprint
  {http://arxiv.org/abs/cond-mat/0409292} {arXiv:cond-mat/0409292} \BibitemShut
  {NoStop}%
\bibitem [{\citenamefont {Schollw{\"o}ck}(2011)}]{schollwock:2011a}%
  \BibitemOpen
  \bibfield  {author} {\bibinfo {author} {\bibfnamefont {U.}~\bibnamefont
  {Schollw{\"o}ck}},\ }\href {\doibase 10.1016/j.aop.2010.09.012} {\bibfield
  {journal} {\bibinfo  {journal} {Ann. Phys.}\ }\textbf {\bibinfo {volume}
  {326}},\ \bibinfo {pages} {96 } (\bibinfo {year} {2011})},\ \Eprint
  {http://arxiv.org/abs/1008.3477} {arXiv:1008.3477} \BibitemShut {NoStop}%
\bibitem [{\citenamefont {Vidal}(2004)}]{vidal:2003a}%
  \BibitemOpen
  \bibfield  {author} {\bibinfo {author} {\bibfnamefont {G.}~\bibnamefont
  {Vidal}},\ }\href {\doibase 10.1103/PhysRevLett.93.040502} {\bibfield
  {journal} {\bibinfo  {journal} {Phys. Rev. Lett.}\ }\textbf {\bibinfo
  {volume} {93}},\ \bibinfo {pages} {040502} (\bibinfo {year} {2004})},\
  \Eprint {http://arxiv.org/abs/quant-ph/0310089} {arXiv:quant-ph/0310089}
  \BibitemShut {NoStop}%
\bibitem [{\citenamefont {Haegeman}\ \emph {et~al.}(2011)\citenamefont
  {Haegeman}, \citenamefont {Cirac}, \citenamefont {Osborne}, \citenamefont
  {Pi\ifmmode~\check{z}\else \v{z}\fi{}orn}, \citenamefont {Verschelde},\ and\
  \citenamefont {Verstraete}}]{haegeman:2011b}%
  \BibitemOpen
  \bibfield  {author} {\bibinfo {author} {\bibfnamefont {J.}~\bibnamefont
  {Haegeman}}, \bibinfo {author} {\bibfnamefont {J.~I.}\ \bibnamefont {Cirac}},
  \bibinfo {author} {\bibfnamefont {T.~J.}\ \bibnamefont {Osborne}}, \bibinfo
  {author} {\bibfnamefont {I.}~\bibnamefont {Pi\ifmmode~\check{z}\else
  \v{z}\fi{}orn}}, \bibinfo {author} {\bibfnamefont {H.}~\bibnamefont
  {Verschelde}}, \ and\ \bibinfo {author} {\bibfnamefont {F.}~\bibnamefont
  {Verstraete}},\ }\href {\doibase 10.1103/PhysRevLett.107.070601} {\bibfield
  {journal} {\bibinfo  {journal} {Phys. Rev. Lett.}\ }\textbf {\bibinfo
  {volume} {107}},\ \bibinfo {pages} {070601} (\bibinfo {year} {2011})},\
  \Eprint {http://arxiv.org/abs/1103.0936} {arXiv:1103.0936} \BibitemShut
  {NoStop}%
\bibitem [{\citenamefont {Corboz}\ and\ \citenamefont
  {Vidal}(2009)}]{corboz:2009a}%
  \BibitemOpen
  \bibfield  {author} {\bibinfo {author} {\bibfnamefont {P.}~\bibnamefont
  {Corboz}}\ and\ \bibinfo {author} {\bibfnamefont {G.}~\bibnamefont {Vidal}},\
  }\href {\doibase 10.1103/PhysRevB.80.165129} {\bibfield  {journal} {\bibinfo
  {journal} {Phys. Rev. B}\ }\textbf {\bibinfo {volume} {80}},\ \bibinfo
  {pages} {165129} (\bibinfo {year} {2009})},\ \Eprint
  {http://arxiv.org/abs/0907.3184} {arXiv:0907.3184} \BibitemShut {NoStop}%
\bibitem [{\citenamefont {Corboz}\ \emph
  {et~al.}(2010{\natexlab{a}})\citenamefont {Corboz}, \citenamefont {Evenbly},
  \citenamefont {Verstraete},\ and\ \citenamefont {Vidal}}]{corboz:2010a}%
  \BibitemOpen
  \bibfield  {author} {\bibinfo {author} {\bibfnamefont {P.}~\bibnamefont
  {Corboz}}, \bibinfo {author} {\bibfnamefont {G.}~\bibnamefont {Evenbly}},
  \bibinfo {author} {\bibfnamefont {F.}~\bibnamefont {Verstraete}}, \ and\
  \bibinfo {author} {\bibfnamefont {G.}~\bibnamefont {Vidal}},\ }\href
  {\doibase 10.1103/PhysRevA.81.010303} {\bibfield  {journal} {\bibinfo
  {journal} {Phys. Rev. A}\ }\textbf {\bibinfo {volume} {81}},\ \bibinfo
  {pages} {010303} (\bibinfo {year} {2010}{\natexlab{a}})},\ \Eprint
  {http://arxiv.org/abs/0904.4151} {arXiv:0904.4151} \BibitemShut {NoStop}%
\bibitem [{\citenamefont {Corboz}\ \emph
  {et~al.}(2010{\natexlab{b}})\citenamefont {Corboz}, \citenamefont {Or\'us},
  \citenamefont {Bauer},\ and\ \citenamefont {Vidal}}]{corboz:2010b}%
  \BibitemOpen
  \bibfield  {author} {\bibinfo {author} {\bibfnamefont {P.}~\bibnamefont
  {Corboz}}, \bibinfo {author} {\bibfnamefont {R.}~\bibnamefont {Or\'us}},
  \bibinfo {author} {\bibfnamefont {B.}~\bibnamefont {Bauer}}, \ and\ \bibinfo
  {author} {\bibfnamefont {G.}~\bibnamefont {Vidal}},\ }\href {\doibase
  10.1103/PhysRevB.81.165104} {\bibfield  {journal} {\bibinfo  {journal} {Phys.
  Rev. B}\ }\textbf {\bibinfo {volume} {81}},\ \bibinfo {pages} {165104}
  (\bibinfo {year} {2010}{\natexlab{b}})},\ \Eprint
  {http://arxiv.org/abs/0912.064} {arXiv:0912.064} \BibitemShut {NoStop}%
\bibitem [{\citenamefont {Kraus}\ \emph {et~al.}(2010)\citenamefont {Kraus},
  \citenamefont {Schuch}, \citenamefont {Verstraete},\ and\ \citenamefont
  {Cirac}}]{kraus:2010a}%
  \BibitemOpen
  \bibfield  {author} {\bibinfo {author} {\bibfnamefont {C.~V.}\ \bibnamefont
  {Kraus}}, \bibinfo {author} {\bibfnamefont {N.}~\bibnamefont {Schuch}},
  \bibinfo {author} {\bibfnamefont {F.}~\bibnamefont {Verstraete}}, \ and\
  \bibinfo {author} {\bibfnamefont {J.~I.}\ \bibnamefont {Cirac}},\ }\href
  {\doibase 10.1103/PhysRevA.81.052338} {\bibfield  {journal} {\bibinfo
  {journal} {Phys. Rev. A}\ }\textbf {\bibinfo {volume} {81}},\ \bibinfo
  {pages} {052338} (\bibinfo {year} {2010})},\ \Eprint
  {http://arxiv.org/abs/0904.4667} {arXiv:0904.4667} \BibitemShut {NoStop}%
\bibitem [{\citenamefont {Haegeman}\ \emph {et~al.}(2012)\citenamefont
  {Haegeman}, \citenamefont {Pirvu}, \citenamefont {Weir}, \citenamefont
  {Cirac}, \citenamefont {Osborne}, \citenamefont {Verschelde},\ and\
  \citenamefont {Verstraete}}]{haegeman:2012a}%
  \BibitemOpen
  \bibfield  {author} {\bibinfo {author} {\bibfnamefont {J.}~\bibnamefont
  {Haegeman}}, \bibinfo {author} {\bibfnamefont {B.}~\bibnamefont {Pirvu}},
  \bibinfo {author} {\bibfnamefont {D.~J.}\ \bibnamefont {Weir}}, \bibinfo
  {author} {\bibfnamefont {J.~I.}\ \bibnamefont {Cirac}}, \bibinfo {author}
  {\bibfnamefont {T.~J.}\ \bibnamefont {Osborne}}, \bibinfo {author}
  {\bibfnamefont {H.}~\bibnamefont {Verschelde}}, \ and\ \bibinfo {author}
  {\bibfnamefont {F.}~\bibnamefont {Verstraete}},\ }\href {\doibase
  10.1103/PhysRevB.85.100408} {\bibfield  {journal} {\bibinfo  {journal} {Phys.
  Rev. B}\ }\textbf {\bibinfo {volume} {85}},\ \bibinfo {pages} {100408}
  (\bibinfo {year} {2012})},\ \Eprint {http://arxiv.org/abs/1103.2286}
  {arXiv:1103.2286} \BibitemShut {NoStop}%
\bibitem [{\citenamefont {Verstraete}\ and\ \citenamefont
  {Cirac}(2004)}]{verstraete:2004a}%
  \BibitemOpen
  \bibfield  {author} {\bibinfo {author} {\bibfnamefont {F.}~\bibnamefont
  {Verstraete}}\ and\ \bibinfo {author} {\bibfnamefont {J.~I.}\ \bibnamefont
  {Cirac}},\ }\href@noop {} {\  (\bibinfo {year} {2004})},\ \Eprint
  {http://arxiv.org/abs/cond-mat/0407066} {arXiv:cond-mat/0407066} \BibitemShut
  {NoStop}%
\bibitem [{\citenamefont {Vidal}(2008)}]{vidal:2006a}%
  \BibitemOpen
  \bibfield  {author} {\bibinfo {author} {\bibfnamefont {G.}~\bibnamefont
  {Vidal}},\ }\href {\doibase 10.1103/PhysRevLett.101.110501} {\bibfield
  {journal} {\bibinfo  {journal} {Phys. Rev. Lett.}\ }\textbf {\bibinfo
  {volume} {101}},\ \bibinfo {pages} {110501} (\bibinfo {year} {2008})},\
  \Eprint {http://arxiv.org/abs/quant-ph/0610099} {arXiv:quant-ph/0610099}
  \BibitemShut {NoStop}%
\bibitem [{\citenamefont {Vidal}(2007)}]{vidal:2007a}%
  \BibitemOpen
  \bibfield  {author} {\bibinfo {author} {\bibfnamefont {G.}~\bibnamefont
  {Vidal}},\ }\href {\doibase 10.1103/PhysRevLett.99.220405} {\bibfield
  {journal} {\bibinfo  {journal} {Phys. Rev. Lett.}\ }\textbf {\bibinfo
  {volume} {99}},\ \bibinfo {pages} {220405} (\bibinfo {year} {2007})},\
  \Eprint {http://arxiv.org/abs/cond-mat/0512165} {arXiv:cond-mat/0512165}
  \BibitemShut {NoStop}%
\bibitem [{\citenamefont {Dalmonte}\ and\ \citenamefont
  {Montangero}(2016)}]{dalmonte_lattice_2016}%
  \BibitemOpen
  \bibfield  {author} {\bibinfo {author} {\bibfnamefont {M.}~\bibnamefont
  {Dalmonte}}\ and\ \bibinfo {author} {\bibfnamefont {S.}~\bibnamefont
  {Montangero}},\ }\href {\doibase 10.1080/00107514.2016.1151199} {\bibfield
  {journal} {\bibinfo  {journal} {Contemp. Phys.}\ }\textbf {\bibinfo {volume}
  {57}},\ \bibinfo {pages} {388} (\bibinfo {year} {2016})},\ \Eprint
  {http://arxiv.org/abs/1602.03776} {arXiv:1602.03776} \BibitemShut {NoStop}%
\bibitem [{\citenamefont {Byrnes}\ \emph {et~al.}(2002)\citenamefont {Byrnes},
  \citenamefont {Sriganesh}, \citenamefont {Bursill},\ and\ \citenamefont
  {Hamer}}]{byrnes:2002a}%
  \BibitemOpen
  \bibfield  {author} {\bibinfo {author} {\bibfnamefont {T.~M.~R.}\
  \bibnamefont {Byrnes}}, \bibinfo {author} {\bibfnamefont {P.}~\bibnamefont
  {Sriganesh}}, \bibinfo {author} {\bibfnamefont {R.~J.}\ \bibnamefont
  {Bursill}}, \ and\ \bibinfo {author} {\bibfnamefont {C.~J.}\ \bibnamefont
  {Hamer}},\ }\href {\doibase 10.1103/PhysRevD.66.013002} {\bibfield  {journal}
  {\bibinfo  {journal} {Phys. Rev. D}\ }\textbf {\bibinfo {volume} {66}},\
  \bibinfo {pages} {013002} (\bibinfo {year} {2002})},\ \Eprint
  {http://arxiv.org/abs/hep-lat/0202014} {arXiv:hep-lat/0202014} \BibitemShut
  {NoStop}%
\bibitem [{\citenamefont {Sugihara}(2005)}]{sugihara:2005a}%
  \BibitemOpen
  \bibfield  {author} {\bibinfo {author} {\bibfnamefont {T.}~\bibnamefont
  {Sugihara}},\ }\href {\doibase 10.1088/1126-6708/2005/07/022} {\bibfield
  {journal} {\bibinfo  {journal} {J. High Energy Phys.}\ }\textbf {\bibinfo
  {volume} {2005}},\ \bibinfo {pages} {22} (\bibinfo {year} {2005})},\ \Eprint
  {http://arxiv.org/abs/hep-lat/0506009} {arXiv:hep-lat/0506009} \BibitemShut
  {NoStop}%
\bibitem [{\citenamefont {Ba{\~n}uls}\ \emph {et~al.}(2013)\citenamefont
  {Ba{\~n}uls}, \citenamefont {Cichy}, \citenamefont {Cirac},\ and\
  \citenamefont {Jansen}}]{banuls:2013a}%
  \BibitemOpen
  \bibfield  {author} {\bibinfo {author} {\bibfnamefont {M.~C.}\ \bibnamefont
  {Ba{\~n}uls}}, \bibinfo {author} {\bibfnamefont {K.}~\bibnamefont {Cichy}},
  \bibinfo {author} {\bibfnamefont {J.~I.}\ \bibnamefont {Cirac}}, \ and\
  \bibinfo {author} {\bibfnamefont {K.}~\bibnamefont {Jansen}},\ }\href
  {\doibase 10.1007/JHEP11(2013)158} {\bibfield  {journal} {\bibinfo  {journal}
  {J. High Energy Phys.}\ } (\bibinfo {year} {2013}),\
  10.1007/JHEP11(2013)158},\ \Eprint {http://arxiv.org/abs/1305.3765}
  {arXiv:1305.3765} \BibitemShut {NoStop}%
\bibitem [{\citenamefont {{Buyens}}\ \emph {et~al.}(2014)\citenamefont
  {{Buyens}}, \citenamefont {{Haegeman}}, \citenamefont {{Van Acoleyen}},
  \citenamefont {{Verschelde}},\ and\ \citenamefont
  {{Verstraete}}}]{buyens:2014}%
  \BibitemOpen
  \bibfield  {author} {\bibinfo {author} {\bibfnamefont {B.}~\bibnamefont
  {{Buyens}}}, \bibinfo {author} {\bibfnamefont {J.}~\bibnamefont
  {{Haegeman}}}, \bibinfo {author} {\bibfnamefont {K.}~\bibnamefont {{Van
  Acoleyen}}}, \bibinfo {author} {\bibfnamefont {H.}~\bibnamefont
  {{Verschelde}}}, \ and\ \bibinfo {author} {\bibfnamefont {F.}~\bibnamefont
  {{Verstraete}}},\ }\href {\doibase 10.1103/PhysRevLett.113.091601} {\bibfield
   {journal} {\bibinfo  {journal} {Phys. Rev. Lett.}\ }\textbf {\bibinfo
  {volume} {113}},\ \bibinfo {pages} {091601} (\bibinfo {year} {2014})},\
  \Eprint {http://arxiv.org/abs/1312.6654} {arXiv:1312.6654} \BibitemShut
  {NoStop}%
\bibitem [{\citenamefont {{Rico}}\ \emph {et~al.}(2014)\citenamefont {{Rico}},
  \citenamefont {{Pichler}}, \citenamefont {{Dalmonte}}, \citenamefont
  {{Zoller}},\ and\ \citenamefont {{Montangero}}}]{rico:2014}%
  \BibitemOpen
  \bibfield  {author} {\bibinfo {author} {\bibfnamefont {E.}~\bibnamefont
  {{Rico}}}, \bibinfo {author} {\bibfnamefont {T.}~\bibnamefont {{Pichler}}},
  \bibinfo {author} {\bibfnamefont {M.}~\bibnamefont {{Dalmonte}}}, \bibinfo
  {author} {\bibfnamefont {P.}~\bibnamefont {{Zoller}}}, \ and\ \bibinfo
  {author} {\bibfnamefont {S.}~\bibnamefont {{Montangero}}},\ }\href {\doibase
  10.1103/PhysRevLett.112.201601} {\bibfield  {journal} {\bibinfo  {journal}
  {Phys. Rev. Lett.}\ }\textbf {\bibinfo {volume} {112}},\ \bibinfo {pages}
  {201601} (\bibinfo {year} {2014})},\ \Eprint {http://arxiv.org/abs/1312.3127}
  {arXiv:1312.3127} \BibitemShut {NoStop}%
\bibitem [{\citenamefont {Pichler}\ \emph {et~al.}(2016)\citenamefont
  {Pichler}, \citenamefont {Dalmonte}, \citenamefont {Rico}, \citenamefont
  {Zoller},\ and\ \citenamefont {Montangero}}]{pichler_real-time_2016}%
  \BibitemOpen
  \bibfield  {author} {\bibinfo {author} {\bibfnamefont {T.}~\bibnamefont
  {Pichler}}, \bibinfo {author} {\bibfnamefont {M.}~\bibnamefont {Dalmonte}},
  \bibinfo {author} {\bibfnamefont {E.}~\bibnamefont {Rico}}, \bibinfo {author}
  {\bibfnamefont {P.}~\bibnamefont {Zoller}}, \ and\ \bibinfo {author}
  {\bibfnamefont {S.}~\bibnamefont {Montangero}},\ }\href {\doibase
  10.1103/PhysRevX.6.011023} {\bibfield  {journal} {\bibinfo  {journal} {Phys.
  Rev. X}\ }\textbf {\bibinfo {volume} {6}},\ \bibinfo {pages} {011023}
  (\bibinfo {year} {2016})}\BibitemShut {NoStop}%
\bibitem [{\citenamefont {Ba{\~n}uls}\ \emph
  {et~al.}(2017{\natexlab{a}})\citenamefont {Ba{\~n}uls}, \citenamefont
  {Cichy}, \citenamefont {Cirac}, \citenamefont {Jansen},\ and\ \citenamefont
  {K{\"u}hn}}]{banuls_density_2017}%
  \BibitemOpen
  \bibfield  {author} {\bibinfo {author} {\bibfnamefont {M.~C.}\ \bibnamefont
  {Ba{\~n}uls}}, \bibinfo {author} {\bibfnamefont {K.}~\bibnamefont {Cichy}},
  \bibinfo {author} {\bibfnamefont {J.~I.}\ \bibnamefont {Cirac}}, \bibinfo
  {author} {\bibfnamefont {K.}~\bibnamefont {Jansen}}, \ and\ \bibinfo {author}
  {\bibfnamefont {S.}~\bibnamefont {K{\"u}hn}},\ }\href {\doibase
  10.1103/PhysRevLett.118.071601} {\bibfield  {journal} {\bibinfo  {journal}
  {Phys. Rev. Lett.}\ }\textbf {\bibinfo {volume} {118}},\ \bibinfo {pages}
  {071601} (\bibinfo {year} {2017}{\natexlab{a}})},\ \Eprint
  {http://arxiv.org/abs/1611.00705} {arXiv:1611.00705} \BibitemShut {NoStop}%
\bibitem [{\citenamefont {Buyens}\ \emph {et~al.}(2017)\citenamefont {Buyens},
  \citenamefont {Montangero}, \citenamefont {Haegeman}, \citenamefont
  {Verstraete},\ and\ \citenamefont
  {Van~Acoleyen}}]{buyens_finite-representation_2017-1}%
  \BibitemOpen
  \bibfield  {author} {\bibinfo {author} {\bibfnamefont {B.}~\bibnamefont
  {Buyens}}, \bibinfo {author} {\bibfnamefont {S.}~\bibnamefont {Montangero}},
  \bibinfo {author} {\bibfnamefont {J.}~\bibnamefont {Haegeman}}, \bibinfo
  {author} {\bibfnamefont {F.}~\bibnamefont {Verstraete}}, \ and\ \bibinfo
  {author} {\bibfnamefont {K.}~\bibnamefont {Van~Acoleyen}},\ }\href {\doibase
  10.1103/PhysRevD.95.094509} {\bibfield  {journal} {\bibinfo  {journal} {Phys.
  Rev. D}\ }\textbf {\bibinfo {volume} {95}},\ \bibinfo {pages} {094509}
  (\bibinfo {year} {2017})},\ \Eprint {http://arxiv.org/abs/1702.08838}
  {arXiv:1702.08838} \BibitemShut {NoStop}%
\bibitem [{\citenamefont {{K{\"u}hn}}\ \emph {et~al.}(2015)\citenamefont
  {{K{\"u}hn}}, \citenamefont {{Zohar}}, \citenamefont {{Cirac}},\ and\
  \citenamefont {{Ba{\~n}uls}}}]{kuhn:2015}%
  \BibitemOpen
  \bibfield  {author} {\bibinfo {author} {\bibfnamefont {S.}~\bibnamefont
  {{K{\"u}hn}}}, \bibinfo {author} {\bibfnamefont {E.}~\bibnamefont {{Zohar}}},
  \bibinfo {author} {\bibfnamefont {J.~I.}\ \bibnamefont {{Cirac}}}, \ and\
  \bibinfo {author} {\bibfnamefont {M.~C.}\ \bibnamefont {{Ba{\~n}uls}}},\
  }\href {\doibase 10.1007/JHEP07(2015)130} {\bibfield  {journal} {\bibinfo
  {journal} {J. High Energ. Phys.}\ }\textbf {\bibinfo {volume} {2015}},\
  \bibinfo {pages} {1} (\bibinfo {year} {2015})},\ \Eprint
  {http://arxiv.org/abs/1505.04441} {arXiv:1505.04441} \BibitemShut {NoStop}%
\bibitem [{\citenamefont {Silvi}\ \emph {et~al.}(2017)\citenamefont {Silvi},
  \citenamefont {Rico}, \citenamefont {Dalmonte}, \citenamefont {Tschirsich},\
  and\ \citenamefont {Montangero}}]{silvi_finite-density_2017}%
  \BibitemOpen
  \bibfield  {author} {\bibinfo {author} {\bibfnamefont {P.}~\bibnamefont
  {Silvi}}, \bibinfo {author} {\bibfnamefont {E.}~\bibnamefont {Rico}},
  \bibinfo {author} {\bibfnamefont {M.}~\bibnamefont {Dalmonte}}, \bibinfo
  {author} {\bibfnamefont {F.}~\bibnamefont {Tschirsich}}, \ and\ \bibinfo
  {author} {\bibfnamefont {S.}~\bibnamefont {Montangero}},\ }\href {\doibase
  10.22331/q-2017-04-25-9} {\bibfield  {journal} {\bibinfo  {journal}
  {Quantum}\ }\textbf {\bibinfo {volume} {1}},\ \bibinfo {pages} {9} (\bibinfo
  {year} {2017})},\ \Eprint {http://arxiv.org/abs/1606.05510}
  {arXiv:1606.05510} \BibitemShut {NoStop}%
\bibitem [{\citenamefont {Ba{\~n}uls}\ \emph
  {et~al.}(2017{\natexlab{b}})\citenamefont {Ba{\~n}uls}, \citenamefont
  {Cichy}, \citenamefont {Cirac}, \citenamefont {Jansen},\ and\ \citenamefont
  {K{\"u}hn}}]{banuls_efficient_2017}%
  \BibitemOpen
  \bibfield  {author} {\bibinfo {author} {\bibfnamefont {M.~C.}\ \bibnamefont
  {Ba{\~n}uls}}, \bibinfo {author} {\bibfnamefont {K.}~\bibnamefont {Cichy}},
  \bibinfo {author} {\bibfnamefont {J.~I.}\ \bibnamefont {Cirac}}, \bibinfo
  {author} {\bibfnamefont {K.}~\bibnamefont {Jansen}}, \ and\ \bibinfo {author}
  {\bibfnamefont {S.}~\bibnamefont {K{\"u}hn}},\ }\href {\doibase
  10.1103/PhysRevX.7.041046} {\bibfield  {journal} {\bibinfo  {journal} {Phys.
  Rev. X}\ }\textbf {\bibinfo {volume} {7}},\ \bibinfo {pages} {041046}
  (\bibinfo {year} {2017}{\natexlab{b}})},\ \Eprint
  {http://arxiv.org/abs/1707.06434} {arXiv:1707.06434} \BibitemShut {NoStop}%
\bibitem [{\citenamefont {Dittrich}\ \emph {et~al.}(2016)\citenamefont
  {Dittrich}, \citenamefont {Mizera},\ and\ \citenamefont
  {Steinhaus}}]{dittrich_decorated_2016}%
  \BibitemOpen
  \bibfield  {author} {\bibinfo {author} {\bibfnamefont {B.}~\bibnamefont
  {Dittrich}}, \bibinfo {author} {\bibfnamefont {S.}~\bibnamefont {Mizera}}, \
  and\ \bibinfo {author} {\bibfnamefont {S.}~\bibnamefont {Steinhaus}},\ }\href
  {\doibase 10.1088/1367-2630/18/5/053009} {\bibfield  {journal} {\bibinfo
  {journal} {New J. Phys.}\ }\textbf {\bibinfo {volume} {18}},\ \bibinfo
  {pages} {053009} (\bibinfo {year} {2016})},\ \Eprint
  {http://arxiv.org/abs/1409.2407} {arXiv:1409.2407} \BibitemShut {NoStop}%
\bibitem [{\citenamefont {Delcamp}\ and\ \citenamefont
  {Dittrich}(2017)}]{delcamp_phase_2017}%
  \BibitemOpen
  \bibfield  {author} {\bibinfo {author} {\bibfnamefont {C.}~\bibnamefont
  {Delcamp}}\ and\ \bibinfo {author} {\bibfnamefont {B.}~\bibnamefont
  {Dittrich}},\ }\href {\doibase 10.1088/1361-6382/aa8f24} {\bibfield
  {journal} {\bibinfo  {journal} {Class. Quantum Grav.}\ }\textbf {\bibinfo
  {volume} {34}},\ \bibinfo {pages} {225006} (\bibinfo {year} {2017})},\
  \Eprint {http://arxiv.org/abs/1612.04506} {arXiv:1612.04506} \BibitemShut
  {NoStop}%
\bibitem [{\citenamefont {Dennis}\ \emph {et~al.}(2002)\citenamefont {Dennis},
  \citenamefont {Kitaev}, \citenamefont {Landahl},\ and\ \citenamefont
  {Preskill}}]{dennis:2002a}%
  \BibitemOpen
  \bibfield  {author} {\bibinfo {author} {\bibfnamefont {E.}~\bibnamefont
  {Dennis}}, \bibinfo {author} {\bibfnamefont {A.}~\bibnamefont {Kitaev}},
  \bibinfo {author} {\bibfnamefont {A.}~\bibnamefont {Landahl}}, \ and\
  \bibinfo {author} {\bibfnamefont {J.}~\bibnamefont {Preskill}},\ }\href
  {\doibase 10.1063/1.1499754} {\bibfield  {journal} {\bibinfo  {journal} {J.
  Math. Phys.}\ }\textbf {\bibinfo {volume} {43}},\ \bibinfo {pages} {4452}
  (\bibinfo {year} {2002})},\ \Eprint {http://arxiv.org/abs/quant-ph/0110143}
  {arXiv:quant-ph/0110143} \BibitemShut {NoStop}%
\bibitem [{\citenamefont {Aguado}\ and\ \citenamefont
  {Vidal}(2008)}]{aguado:2008a}%
  \BibitemOpen
  \bibfield  {author} {\bibinfo {author} {\bibfnamefont {M.}~\bibnamefont
  {Aguado}}\ and\ \bibinfo {author} {\bibfnamefont {G.}~\bibnamefont {Vidal}},\
  }\href {\doibase 10.1103/PhysRevLett.100.070404} {\bibfield  {journal}
  {\bibinfo  {journal} {Phys. Rev. Lett.}\ }\textbf {\bibinfo {volume} {100}},\
  \bibinfo {pages} {070404} (\bibinfo {year} {2008})},\ \Eprint
  {http://arxiv.org/abs/0712.0348} {arXiv:0712.0348} \BibitemShut {NoStop}%
\bibitem [{\citenamefont {Buerschaper}\ \emph {et~al.}(2009)\citenamefont
  {Buerschaper}, \citenamefont {Aguado},\ and\ \citenamefont
  {Vidal}}]{buerschaper:2009a}%
  \BibitemOpen
  \bibfield  {author} {\bibinfo {author} {\bibfnamefont {O.}~\bibnamefont
  {Buerschaper}}, \bibinfo {author} {\bibfnamefont {M.}~\bibnamefont {Aguado}},
  \ and\ \bibinfo {author} {\bibfnamefont {G.}~\bibnamefont {Vidal}},\ }\href
  {\doibase 10.1103/PhysRevB.79.085119} {\bibfield  {journal} {\bibinfo
  {journal} {Phys. Rev. B}\ }\textbf {\bibinfo {volume} {79}},\ \bibinfo
  {pages} {085119} (\bibinfo {year} {2009})},\ \Eprint
  {http://arxiv.org/abs/0809.2393} {arXiv:0809.2393} \BibitemShut {NoStop}%
\bibitem [{\citenamefont {K\"onig}\ \emph {et~al.}(2009)\citenamefont
  {K\"onig}, \citenamefont {Reichardt},\ and\ \citenamefont
  {Vidal}}]{koenig:2009a}%
  \BibitemOpen
  \bibfield  {author} {\bibinfo {author} {\bibfnamefont {R.}~\bibnamefont
  {K\"onig}}, \bibinfo {author} {\bibfnamefont {B.~W.}\ \bibnamefont
  {Reichardt}}, \ and\ \bibinfo {author} {\bibfnamefont {G.}~\bibnamefont
  {Vidal}},\ }\href {\doibase 10.1103/PhysRevB.79.195123} {\bibfield  {journal}
  {\bibinfo  {journal} {Phys. Rev. B}\ }\textbf {\bibinfo {volume} {79}},\
  \bibinfo {pages} {195123} (\bibinfo {year} {2009})},\ \Eprint
  {http://arxiv.org/abs/0806.4583} {arXiv:0806.4583} \BibitemShut {NoStop}%
\bibitem [{\citenamefont {Tagliacozzo}\ and\ \citenamefont
  {Vidal}(2011)}]{tagliacozzo:2011a}%
  \BibitemOpen
  \bibfield  {author} {\bibinfo {author} {\bibfnamefont {L.}~\bibnamefont
  {Tagliacozzo}}\ and\ \bibinfo {author} {\bibfnamefont {G.}~\bibnamefont
  {Vidal}},\ }\href {\doibase 10.1103/PhysRevB.83.115127} {\bibfield  {journal}
  {\bibinfo  {journal} {Phys. Rev. B}\ }\textbf {\bibinfo {volume} {83}},\
  \bibinfo {pages} {115127} (\bibinfo {year} {2011})},\ \Eprint
  {http://arxiv.org/abs/1007.4145} {arXiv:1007.4145} \BibitemShut {NoStop}%
\bibitem [{\citenamefont {Tagliacozzo}\ \emph {et~al.}(2014)\citenamefont
  {Tagliacozzo}, \citenamefont {Celi},\ and\ \citenamefont
  {Lewenstein}}]{tagliacozzo_tensor_2014}%
  \BibitemOpen
  \bibfield  {author} {\bibinfo {author} {\bibfnamefont {L.}~\bibnamefont
  {Tagliacozzo}}, \bibinfo {author} {\bibfnamefont {A.}~\bibnamefont {Celi}}, \
  and\ \bibinfo {author} {\bibfnamefont {M.}~\bibnamefont {Lewenstein}},\
  }\href {\doibase 10.1103/PhysRevX.4.041024} {\bibfield  {journal} {\bibinfo
  {journal} {Phys. Rev. X}\ }\textbf {\bibinfo {volume} {4}},\ \bibinfo {pages}
  {041024} (\bibinfo {year} {2014})},\ \Eprint {http://arxiv.org/abs/1405.4811}
  {arXiv:1405.4811} \BibitemShut {NoStop}%
\bibitem [{\citenamefont {Silvi}\ \emph {et~al.}(2014)\citenamefont {Silvi},
  \citenamefont {Rico}, \citenamefont {Calarco},\ and\ \citenamefont
  {Montangero}}]{silvi_lattice_2014}%
  \BibitemOpen
  \bibfield  {author} {\bibinfo {author} {\bibfnamefont {P.}~\bibnamefont
  {Silvi}}, \bibinfo {author} {\bibfnamefont {E.}~\bibnamefont {Rico}},
  \bibinfo {author} {\bibfnamefont {T.}~\bibnamefont {Calarco}}, \ and\
  \bibinfo {author} {\bibfnamefont {S.}~\bibnamefont {Montangero}},\ }\href
  {\doibase 10.1088/1367-2630/16/10/103015} {\bibfield  {journal} {\bibinfo
  {journal} {New J. Phys.}\ }\textbf {\bibinfo {volume} {16}},\ \bibinfo
  {pages} {103015} (\bibinfo {year} {2014})},\ \Eprint
  {http://arxiv.org/abs/1404.7439} {arXiv:1404.7439} \BibitemShut {NoStop}%
\bibitem [{\citenamefont {Zohar}\ \emph {et~al.}(2015)\citenamefont {Zohar},
  \citenamefont {Burrello}, \citenamefont {Wahl},\ and\ \citenamefont
  {Cirac}}]{zohar:2015}%
  \BibitemOpen
  \bibfield  {author} {\bibinfo {author} {\bibfnamefont {E.}~\bibnamefont
  {Zohar}}, \bibinfo {author} {\bibfnamefont {M.}~\bibnamefont {Burrello}},
  \bibinfo {author} {\bibfnamefont {T.~B.}\ \bibnamefont {Wahl}}, \ and\
  \bibinfo {author} {\bibfnamefont {J.~I.}\ \bibnamefont {Cirac}},\ }\href
  {\doibase 10.1016/j.aop.2015.10.009} {\bibfield  {journal} {\bibinfo
  {journal} {Ann. Phys.}\ }\textbf {\bibinfo {volume} {363}},\ \bibinfo {pages}
  {385} (\bibinfo {year} {2015})},\ \Eprint {http://arxiv.org/abs/1507.08837}
  {arXiv:1507.08837} \BibitemShut {NoStop}%
\bibitem [{\citenamefont {Zohar}\ and\ \citenamefont
  {Burrello}(2016)}]{zohar:2016}%
  \BibitemOpen
  \bibfield  {author} {\bibinfo {author} {\bibfnamefont {E.}~\bibnamefont
  {Zohar}}\ and\ \bibinfo {author} {\bibfnamefont {M.}~\bibnamefont
  {Burrello}},\ }\href {\doibase 10.1088/1367-2630/18/4/043008} {\bibfield
  {journal} {\bibinfo  {journal} {New J. Phys.}\ }\textbf {\bibinfo {volume}
  {18}},\ \bibinfo {pages} {043008} (\bibinfo {year} {2016})},\ \Eprint
  {http://arxiv.org/abs/1511.08426} {arXiv:1511.08426} \BibitemShut {NoStop}%
\bibitem [{\citenamefont {Zohar}\ \emph {et~al.}(2016)\citenamefont {Zohar},
  \citenamefont {Wahl}, \citenamefont {Burrello},\ and\ \citenamefont
  {Cirac}}]{zohar_projected_2016}%
  \BibitemOpen
  \bibfield  {author} {\bibinfo {author} {\bibfnamefont {E.}~\bibnamefont
  {Zohar}}, \bibinfo {author} {\bibfnamefont {T.~B.}\ \bibnamefont {Wahl}},
  \bibinfo {author} {\bibfnamefont {M.}~\bibnamefont {Burrello}}, \ and\
  \bibinfo {author} {\bibfnamefont {J.~I.}\ \bibnamefont {Cirac}},\ }\href
  {\doibase 10.1016/j.aop.2016.08.008} {\bibfield  {journal} {\bibinfo
  {journal} {Ann. Phys.}\ }\textbf {\bibinfo {volume} {374}},\ \bibinfo {pages}
  {84} (\bibinfo {year} {2016})},\ \Eprint {http://arxiv.org/abs/1607.08115}
  {arXiv:1607.08115} \BibitemShut {NoStop}%
\bibitem [{\citenamefont {Kull}\ \emph {et~al.}(2017)\citenamefont {Kull},
  \citenamefont {Molnar}, \citenamefont {Zohar},\ and\ \citenamefont
  {Cirac}}]{kull_classification_2017}%
  \BibitemOpen
  \bibfield  {author} {\bibinfo {author} {\bibfnamefont {I.}~\bibnamefont
  {Kull}}, \bibinfo {author} {\bibfnamefont {A.}~\bibnamefont {Molnar}},
  \bibinfo {author} {\bibfnamefont {E.}~\bibnamefont {Zohar}}, \ and\ \bibinfo
  {author} {\bibfnamefont {J.~I.}\ \bibnamefont {Cirac}},\ }\href {\doibase
  10.1016/j.aop.2017.08.029} {\bibfield  {journal} {\bibinfo  {journal} {Ann.
  Phys.}\ }\textbf {\bibinfo {volume} {386}},\ \bibinfo {pages} {199} (\bibinfo
  {year} {2017})},\ \Eprint {http://arxiv.org/abs/1708.00362}
  {arXiv:1708.00362} \BibitemShut {NoStop}%
\bibitem [{\citenamefont {Haegeman}\ \emph {et~al.}(2013)\citenamefont
  {Haegeman}, \citenamefont {Osborne}, \citenamefont {Verschelde},\ and\
  \citenamefont {Verstraete}}]{haegeman:2011a}%
  \BibitemOpen
  \bibfield  {author} {\bibinfo {author} {\bibfnamefont {J.}~\bibnamefont
  {Haegeman}}, \bibinfo {author} {\bibfnamefont {T.~J.}\ \bibnamefont
  {Osborne}}, \bibinfo {author} {\bibfnamefont {H.}~\bibnamefont {Verschelde}},
  \ and\ \bibinfo {author} {\bibfnamefont {F.}~\bibnamefont {Verstraete}},\
  }\href {http://arxiv.org/abs/1102.5524} {\bibfield  {journal} {\bibinfo
  {journal} {Phys. Rev. Lett.}\ }\textbf {\bibinfo {volume} {110}},\ \bibinfo
  {pages} {100402} (\bibinfo {year} {2013})},\ \Eprint
  {http://arxiv.org/abs/1102.5524} {arXiv:1102.5524} \BibitemShut {NoStop}%
\bibitem [{\citenamefont {Kogut}\ and\ \citenamefont
  {Susskind}(1975)}]{kogut:1975a}%
  \BibitemOpen
  \bibfield  {author} {\bibinfo {author} {\bibfnamefont {J.}~\bibnamefont
  {Kogut}}\ and\ \bibinfo {author} {\bibfnamefont {L.}~\bibnamefont
  {Susskind}},\ }\href {\doibase 10.1103/PhysRevD.11.395} {\bibfield  {journal}
  {\bibinfo  {journal} {Phys. Rev. D}\ }\textbf {\bibinfo {volume} {11}},\
  \bibinfo {pages} {395} (\bibinfo {year} {1975})}\BibitemShut {NoStop}%
\bibitem [{\citenamefont {Wilson}(1975)}]{wilson:1975a}%
  \BibitemOpen
  \bibfield  {author} {\bibinfo {author} {\bibfnamefont {K.~G.}\ \bibnamefont
  {Wilson}},\ }\href {\doibase 10.1103/RevModPhys.47.773} {\bibfield  {journal}
  {\bibinfo  {journal} {Rev. Modern Phys.}\ }\textbf {\bibinfo {volume} {47}},\
  \bibinfo {pages} {773} (\bibinfo {year} {1975})}\BibitemShut {NoStop}%
\bibitem [{\citenamefont {Ba{\l}aban}(1985{\natexlab{a}})}]{balaban:1985a}%
  \BibitemOpen
  \bibfield  {author} {\bibinfo {author} {\bibfnamefont {T.}~\bibnamefont
  {Ba{\l}aban}},\ }\href {http://projecteuclid.org/euclid.cmp/1103942283}
  {\bibfield  {journal} {\bibinfo  {journal} {Comm. Math. Phys.}\ }\textbf
  {\bibinfo {volume} {98}},\ \bibinfo {pages} {17} (\bibinfo {year}
  {1985}{\natexlab{a}})}\BibitemShut {NoStop}%
\bibitem [{\citenamefont {Ba{\l}aban}(1988{\natexlab{a}})}]{balaban:1988a}%
  \BibitemOpen
  \bibfield  {author} {\bibinfo {author} {\bibfnamefont {T.}~\bibnamefont
  {Ba{\l}aban}},\ }\href {http://projecteuclid.org/euclid.cmp/1104162401}
  {\bibfield  {journal} {\bibinfo  {journal} {Comm. Math. Phys.}\ }\textbf
  {\bibinfo {volume} {119}},\ \bibinfo {pages} {243} (\bibinfo {year}
  {1988}{\natexlab{a}})}\BibitemShut {NoStop}%
\bibitem [{\citenamefont {Ba{\l}aban}(1984{\natexlab{a}})}]{balaban:1984a}%
  \BibitemOpen
  \bibfield  {author} {\bibinfo {author} {\bibfnamefont {T.}~\bibnamefont
  {Ba{\l}aban}},\ }\href {http://projecteuclid.org/euclid.cmp/1103941451}
  {\bibfield  {journal} {\bibinfo  {journal} {Comm. Math. Phys.}\ }\textbf
  {\bibinfo {volume} {95}},\ \bibinfo {pages} {17} (\bibinfo {year}
  {1984}{\natexlab{a}})}\BibitemShut {NoStop}%
\bibitem [{\citenamefont {Ba{\l}aban}(1984{\natexlab{b}})}]{balaban:1984b}%
  \BibitemOpen
  \bibfield  {author} {\bibinfo {author} {\bibfnamefont {T.}~\bibnamefont
  {Ba{\l}aban}},\ }\href {http://projecteuclid.org/euclid.cmp/1103941783}
  {\bibfield  {journal} {\bibinfo  {journal} {Comm. Math. Phys.}\ }\textbf
  {\bibinfo {volume} {96}},\ \bibinfo {pages} {223} (\bibinfo {year}
  {1984}{\natexlab{b}})}\BibitemShut {NoStop}%
\bibitem [{\citenamefont {Ba{\l}aban}(1985{\natexlab{b}})}]{balaban:1985b}%
  \BibitemOpen
  \bibfield  {author} {\bibinfo {author} {\bibfnamefont {T.}~\bibnamefont
  {Ba{\l}aban}},\ }\href {http://projecteuclid.org/euclid.cmp/1103942769}
  {\bibfield  {journal} {\bibinfo  {journal} {Comm. Math. Phys.}\ }\textbf
  {\bibinfo {volume} {99}},\ \bibinfo {pages} {389} (\bibinfo {year}
  {1985}{\natexlab{b}})}\BibitemShut {NoStop}%
\bibitem [{\citenamefont {Ba{\l}aban}(1985{\natexlab{c}})}]{balaban:1985c}%
  \BibitemOpen
  \bibfield  {author} {\bibinfo {author} {\bibfnamefont {T.}~\bibnamefont
  {Ba{\l}aban}},\ }\href {http://projecteuclid.org/euclid.cmp/1103942611}
  {\bibfield  {journal} {\bibinfo  {journal} {Comm. Math. Phys.}\ }\textbf
  {\bibinfo {volume} {99}},\ \bibinfo {pages} {75} (\bibinfo {year}
  {1985}{\natexlab{c}})}\BibitemShut {NoStop}%
\bibitem [{\citenamefont {Ba{\l}aban}(1985{\natexlab{d}})}]{balaban:1985d}%
  \BibitemOpen
  \bibfield  {author} {\bibinfo {author} {\bibfnamefont {T.}~\bibnamefont
  {Ba{\l}aban}},\ }\href {http://projecteuclid.org/euclid.cmp/1104114382}
  {\bibfield  {journal} {\bibinfo  {journal} {Comm. Math. Phys.}\ }\textbf
  {\bibinfo {volume} {102}},\ \bibinfo {pages} {255} (\bibinfo {year}
  {1985}{\natexlab{d}})}\BibitemShut {NoStop}%
\bibitem [{\citenamefont {Ba{\l}aban}(1989{\natexlab{a}})}]{balaban:1989a}%
  \BibitemOpen
  \bibfield  {author} {\bibinfo {author} {\bibfnamefont {T.}~\bibnamefont
  {Ba{\l}aban}},\ }\href {http://projecteuclid.org/euclid.cmp/1104178393}
  {\bibfield  {journal} {\bibinfo  {journal} {Comm. Math. Phys.}\ }\textbf
  {\bibinfo {volume} {122}},\ \bibinfo {pages} {175} (\bibinfo {year}
  {1989}{\natexlab{a}})}\BibitemShut {NoStop}%
\bibitem [{\citenamefont {Ba{\l}aban}(1989{\natexlab{b}})}]{balaban:1989b}%
  \BibitemOpen
  \bibfield  {author} {\bibinfo {author} {\bibfnamefont {T.}~\bibnamefont
  {Ba{\l}aban}},\ }\href {http://projecteuclid.org/euclid.cmp/1104178467}
  {\bibfield  {journal} {\bibinfo  {journal} {Comm. Math. Phys.}\ }\textbf
  {\bibinfo {volume} {122}},\ \bibinfo {pages} {355} (\bibinfo {year}
  {1989}{\natexlab{b}})}\BibitemShut {NoStop}%
\bibitem [{\citenamefont {Ba{\l}aban}(1987)}]{balaban:1987a}%
  \BibitemOpen
  \bibfield  {author} {\bibinfo {author} {\bibfnamefont {T.}~\bibnamefont
  {Ba{\l}aban}},\ }\href {http://projecteuclid.org/euclid.cmp/1104116842}
  {\bibfield  {journal} {\bibinfo  {journal} {Comm. Math. Phys.}\ }\textbf
  {\bibinfo {volume} {109}},\ \bibinfo {pages} {249} (\bibinfo {year}
  {1987})}\BibitemShut {NoStop}%
\bibitem [{\citenamefont {Ba{\l}aban}(1988{\natexlab{b}})}]{balaban:1988b}%
  \BibitemOpen
  \bibfield  {author} {\bibinfo {author} {\bibfnamefont {T.}~\bibnamefont
  {Ba{\l}aban}},\ }\href {http://projecteuclid.org/euclid.cmp/1104161193}
  {\bibfield  {journal} {\bibinfo  {journal} {Comm. Math. Phys.}\ }\textbf
  {\bibinfo {volume} {116}},\ \bibinfo {pages} {1} (\bibinfo {year}
  {1988}{\natexlab{b}})}\BibitemShut {NoStop}%
\bibitem [{\citenamefont {Federbush}(1987{\natexlab{a}})}]{federbush:1987a}%
  \BibitemOpen
  \bibfield  {author} {\bibinfo {author} {\bibfnamefont {P.}~\bibnamefont
  {Federbush}},\ }\href {http://www.numdam.org/item?id=AIHPA_1987__47_1_17_0}
  {\bibfield  {journal} {\bibinfo  {journal} {Ann. Inst. Henri Poincar\'e}\
  }\textbf {\bibinfo {volume} {47}},\ \bibinfo {pages} {17} (\bibinfo {year}
  {1987}{\natexlab{a}})}\BibitemShut {NoStop}%
\bibitem [{\citenamefont {Federbush}(1986)}]{federbush:1986a}%
  \BibitemOpen
  \bibfield  {author} {\bibinfo {author} {\bibfnamefont {P.}~\bibnamefont
  {Federbush}},\ }\href {http://projecteuclid.org/euclid.cmp/1104116026}
  {\bibfield  {journal} {\bibinfo  {journal} {Comm. Math. Phys.}\ }\textbf
  {\bibinfo {volume} {107}},\ \bibinfo {pages} {319} (\bibinfo {year}
  {1986})}\BibitemShut {NoStop}%
\bibitem [{\citenamefont {Federbush}\ and\ \citenamefont
  {Williamson}(1987)}]{federbush:1987b}%
  \BibitemOpen
  \bibfield  {author} {\bibinfo {author} {\bibfnamefont {P.}~\bibnamefont
  {Federbush}}\ and\ \bibinfo {author} {\bibfnamefont {C.}~\bibnamefont
  {Williamson}},\ }\href {\doibase doi:10.1063/1.527495} {\bibfield  {journal}
  {\bibinfo  {journal} {J. Math. Phys.}\ }\textbf {\bibinfo {volume} {28}},\
  \bibinfo {pages} {1416} (\bibinfo {year} {1987})}\BibitemShut {NoStop}%
\bibitem [{\citenamefont {Federbush}(1987{\natexlab{b}})}]{federbush:1987c}%
  \BibitemOpen
  \bibfield  {author} {\bibinfo {author} {\bibfnamefont {P.}~\bibnamefont
  {Federbush}},\ }\href {http://projecteuclid.org/euclid.cmp/1104159240}
  {\bibfield  {journal} {\bibinfo  {journal} {Comm. Math. Phys.}\ }\textbf
  {\bibinfo {volume} {110}},\ \bibinfo {pages} {293} (\bibinfo {year}
  {1987}{\natexlab{b}})}\BibitemShut {NoStop}%
\bibitem [{\citenamefont {Federbush}(1988)}]{federbush:1988a}%
  \BibitemOpen
  \bibfield  {author} {\bibinfo {author} {\bibfnamefont {P.}~\bibnamefont
  {Federbush}},\ }\href {http://projecteuclid.org/euclid.cmp/1104160585}
  {\bibfield  {journal} {\bibinfo  {journal} {Comm. Math. Phys.}\ }\textbf
  {\bibinfo {volume} {114}},\ \bibinfo {pages} {317} (\bibinfo {year}
  {1988})}\BibitemShut {NoStop}%
\bibitem [{\citenamefont {Federbush}(1990)}]{federbush:1990a}%
  \BibitemOpen
  \bibfield  {author} {\bibinfo {author} {\bibfnamefont {P.}~\bibnamefont
  {Federbush}},\ }\href {http://projecteuclid.org/euclid.cmp/1104180215}
  {\bibfield  {journal} {\bibinfo  {journal} {Comm. Math. Phys.}\ }\textbf
  {\bibinfo {volume} {127}},\ \bibinfo {pages} {433} (\bibinfo {year}
  {1990})}\BibitemShut {NoStop}%
\bibitem [{\citenamefont {Magnen}\ \emph {et~al.}(1993)\citenamefont {Magnen},
  \citenamefont {Rivasseau},\ and\ \citenamefont {S\'en\'eor}}]{magnen:1993a}%
  \BibitemOpen
  \bibfield  {author} {\bibinfo {author} {\bibfnamefont {J.}~\bibnamefont
  {Magnen}}, \bibinfo {author} {\bibfnamefont {V.}~\bibnamefont {Rivasseau}}, \
  and\ \bibinfo {author} {\bibfnamefont {R.}~\bibnamefont {S\'en\'eor}},\
  }\href {\doibase 10.1007/BF02097397} {\bibfield  {journal} {\bibinfo
  {journal} {Comm. Math. Phys.}\ }\textbf {\bibinfo {volume} {155}},\ \bibinfo
  {pages} {325} (\bibinfo {year} {1993})}\BibitemShut {NoStop}%
\bibitem [{Note2()}]{Note2}%
  \BibitemOpen
  \bibinfo {note} {If we are to speculate wildly for a moment we would say that
  an efficiently contractible ansatz for the ground state of Yang-Mills could
  well lead to progress on question of existence in the constructive QFT sense:
  one only needs to look to condensed matter physics and the examples of the
  BCS state and the Laughlin wavefunctions to see how a good ansatz leads to
  both physical and mathematical progress.}\BibitemShut {Stop}%
\bibitem [{\citenamefont {Baez}(1996)}]{baez:1996a}%
  \BibitemOpen
  \bibfield  {author} {\bibinfo {author} {\bibfnamefont {J.~C.}\ \bibnamefont
  {Baez}},\ }\href {\doibase 10.1006/aima.1996.0012} {\bibfield  {journal}
  {\bibinfo  {journal} {Adv. Math.}\ }\textbf {\bibinfo {volume} {117}},\
  \bibinfo {pages} {253} (\bibinfo {year} {1996})},\ \Eprint
  {http://arxiv.org/abs/gr-qc/9411007} {arXiv:gr-qc/9411007} \BibitemShut
  {NoStop}%
\bibitem [{Note3()}]{Note3}%
  \BibitemOpen
  \bibinfo {note} {The (conjectured) fundamental excitation of pure Yang-Mills
  theory, the \protect \emph {glueball}, has not yet been observed
  directly.}\BibitemShut {Stop}%
\bibitem [{Note4()}]{Note4}%
  \BibitemOpen
  \bibinfo {note} {Presumably $\Delta (g_H)$ could be monotonically decreasing
  as a function of $g_H$, but this is unlikely as the addition of irrelevant
  ultraviolet interactions could easily modify the behaviour of the gap for
  large coupling $g_H$ without leading to any impact on the large-scale
  physics.}\BibitemShut {Stop}%
\bibitem [{\citenamefont {Seiler}()}]{seiler:2003a}%
  \BibitemOpen
  \bibfield  {author} {\bibinfo {author} {\bibfnamefont {E.}~\bibnamefont
  {Seiler}},\ }\href@noop {} {\enquote {\bibinfo {title} {The case against
  asymptotic freedom},}\ }\Eprint {http://arxiv.org/abs/hep-th/0312015}
  {arXiv:hep-th/0312015} \BibitemShut {NoStop}%
\bibitem [{cmp()}]{cmproblem:1st}%
  \BibitemOpen
  \href {{h}ttp://www.claymath.org/millennium-problems} {}\bibinfo {note}
  {{h}ttp://www.claymath.org/millennium-problems}\BibitemShut {NoStop}%
\bibitem [{\citenamefont {Polyakov}(1975)}]{polyakov:1975a}%
  \BibitemOpen
  \bibfield  {author} {\bibinfo {author} {\bibfnamefont {A.~M.}\ \bibnamefont
  {Polyakov}},\ }\href {\doibase 10.1016/0370-2693(75)90162-8} {\bibfield
  {journal} {\bibinfo  {journal} {Phys. Lett. B}\ }\textbf {\bibinfo {volume}
  {59}},\ \bibinfo {pages} {82 } (\bibinfo {year} {1975})}\BibitemShut
  {NoStop}%
\bibitem [{\citenamefont {Villain}(975a)}]{villain:1975a}%
  \BibitemOpen
  \bibfield  {author} {\bibinfo {author} {\bibfnamefont {J.}~\bibnamefont
  {Villain}},\ }\href {\doibase 10.1051/jphys:01975003606058100} {\bibfield
  {journal} {\bibinfo  {journal} {J. Phys. France}\ }\textbf {\bibinfo {volume}
  {36}},\ \bibinfo {pages} {581 } (\bibinfo {year} {1975a})}\BibitemShut
  {NoStop}%
\bibitem [{\citenamefont {Polyakov}(1977)}]{polyakov:1977a}%
  \BibitemOpen
  \bibfield  {author} {\bibinfo {author} {\bibfnamefont {A.~M.}\ \bibnamefont
  {Polyakov}},\ }\href {\doibase 10.1016/0550-3213(77)90086-4} {\bibfield
  {journal} {\bibinfo  {journal} {Nucl. Phys. B}\ }\textbf {\bibinfo {volume}
  {120}},\ \bibinfo {pages} {429 } (\bibinfo {year} {1977})}\BibitemShut
  {NoStop}%
\bibitem [{\citenamefont {Banks}\ \emph {et~al.}(1977)\citenamefont {Banks},
  \citenamefont {Myerson},\ and\ \citenamefont {Kogut}}]{banks:1977a}%
  \BibitemOpen
  \bibfield  {author} {\bibinfo {author} {\bibfnamefont {T.}~\bibnamefont
  {Banks}}, \bibinfo {author} {\bibfnamefont {R.}~\bibnamefont {Myerson}}, \
  and\ \bibinfo {author} {\bibfnamefont {J.}~\bibnamefont {Kogut}},\ }\href
  {\doibase 10.1016/0550-3213(77)90129-8} {\bibfield  {journal} {\bibinfo
  {journal} {Nucl. Phys. B}\ }\textbf {\bibinfo {volume} {129}},\ \bibinfo
  {pages} {493 } (\bibinfo {year} {1977})}\BibitemShut {NoStop}%
\bibitem [{\citenamefont {Peskin}(1978)}]{peskin:1978a}%
  \BibitemOpen
  \bibfield  {author} {\bibinfo {author} {\bibfnamefont {M.~E.}\ \bibnamefont
  {Peskin}},\ }\href {\doibase 10.1016/0003-4916(78)90252-X} {\bibfield
  {journal} {\bibinfo  {journal} {Ann. Phys.}\ }\textbf {\bibinfo {volume}
  {113}},\ \bibinfo {pages} {122 } (\bibinfo {year} {1978})}\BibitemShut
  {NoStop}%
\bibitem [{\citenamefont {Kogut}\ and\ \citenamefont
  {Stephanov}(2004)}]{kogut:2004a}%
  \BibitemOpen
  \bibfield  {author} {\bibinfo {author} {\bibfnamefont {J.~B.}\ \bibnamefont
  {Kogut}}\ and\ \bibinfo {author} {\bibfnamefont {M.~A.}\ \bibnamefont
  {Stephanov}},\ }\href@noop {} {\emph {\bibinfo {title} {The {P}hases of
  {Q}uantum {C}hromodynamics: {F}rom {C}onfinement to {E}xtreme
  {E}nvironments}}},\ \bibinfo {series} {Cambridge monographs on particle
  physics, nuclear physics, and cosmology}, Vol.~\bibinfo {volume} {21}\
  (\bibinfo  {publisher} {Cambridge University Press},\ \bibinfo {address}
  {Cambridge},\ \bibinfo {year} {2004})\BibitemShut {NoStop}%
\bibitem [{\citenamefont {Hatcher}(2002)}]{hatcher:2002}%
  \BibitemOpen
  \bibfield  {author} {\bibinfo {author} {\bibfnamefont {A.}~\bibnamefont
  {Hatcher}},\ }\href@noop {} {\emph {\bibinfo {title} {Algebraic
  {T}opology}}}\ (\bibinfo  {publisher} {Cambridge University Press},\ \bibinfo
  {year} {2002})\BibitemShut {NoStop}%
\bibitem [{\citenamefont {Ligterink}\ \emph {et~al.}(2000)\citenamefont
  {Ligterink}, \citenamefont {Walet},\ and\ \citenamefont
  {Bishop}}]{ligterink:2000a}%
  \BibitemOpen
  \bibfield  {author} {\bibinfo {author} {\bibfnamefont {N.~E.}\ \bibnamefont
  {Ligterink}}, \bibinfo {author} {\bibfnamefont {N.~R.}\ \bibnamefont
  {Walet}}, \ and\ \bibinfo {author} {\bibfnamefont {R.~F.}\ \bibnamefont
  {Bishop}},\ }\href {\doibase 10.1006/aphy.2000.6070} {\bibfield  {journal}
  {\bibinfo  {journal} {Ann. Phys.}\ }\textbf {\bibinfo {volume} {284}},\
  \bibinfo {pages} {215 } (\bibinfo {year} {2000})},\ \Eprint
  {http://arxiv.org/abs/hep-lat/0001028} {arXiv:hep-lat/0001028} \BibitemShut
  {NoStop}%
\bibitem [{\citenamefont {Endres}\ \emph {et~al.}(2015)\citenamefont {Endres},
  \citenamefont {Brower}, \citenamefont {Detmold}, \citenamefont {Orginos},\
  and\ \citenamefont {Pochinsky}}]{endres:2015}%
  \BibitemOpen
  \bibfield  {author} {\bibinfo {author} {\bibfnamefont {M.~G.}\ \bibnamefont
  {Endres}}, \bibinfo {author} {\bibfnamefont {R.~C.}\ \bibnamefont {Brower}},
  \bibinfo {author} {\bibfnamefont {W.}~\bibnamefont {Detmold}}, \bibinfo
  {author} {\bibfnamefont {K.}~\bibnamefont {Orginos}}, \ and\ \bibinfo
  {author} {\bibfnamefont {A.~V.}\ \bibnamefont {Pochinsky}},\ }\href {\doibase
  10.1103/PhysRevD.92.114516} {\bibfield  {journal} {\bibinfo  {journal} {Phys.
  Rev. D}\ }\textbf {\bibinfo {volume} {92}},\ \bibinfo {pages} {114516}
  (\bibinfo {year} {2015})},\ \Eprint {http://arxiv.org/abs/1510.04675}
  {arXiv:1510.04675} \BibitemShut {NoStop}%
\bibitem [{\citenamefont {Shoemake}(1985)}]{shoemake:1985a}%
  \BibitemOpen
  \bibfield  {author} {\bibinfo {author} {\bibfnamefont {K.}~\bibnamefont
  {Shoemake}},\ }\href {\doibase 10.1145/325334.325242} {\bibfield  {journal}
  {\bibinfo  {journal} {ACM SIGGRAPH computer graphics}\ }\textbf {\bibinfo
  {volume} {19}},\ \bibinfo {pages} {245} (\bibinfo {year} {1985})}\BibitemShut
  {NoStop}%
\bibitem [{\citenamefont {L\"uscher}(2010{\natexlab{a}})}]{luscher:2010a}%
  \BibitemOpen
  \bibfield  {author} {\bibinfo {author} {\bibfnamefont {M.}~\bibnamefont
  {L\"uscher}},\ }\href {\doibase 10.1007/s00220-009-0953-7} {\bibfield
  {journal} {\bibinfo  {journal} {Comm. Math. Phys.}\ }\textbf {\bibinfo
  {volume} {293}},\  (\bibinfo {year} {2010}{\natexlab{a}})},\ \Eprint
  {http://arxiv.org/abs/0907.5491} {arXiv:0907.5491} \BibitemShut {NoStop}%
\bibitem [{\citenamefont {L\"uscher}(2010{\natexlab{b}})}]{luscher:2010b}%
  \BibitemOpen
  \bibfield  {author} {\bibinfo {author} {\bibfnamefont {M.}~\bibnamefont
  {L\"uscher}},\ }\href {\doibase 10.1007/JHEP08(2010)071} {\bibfield
  {journal} {\bibinfo  {journal} {J. High Energy Phys.}\ }\textbf {\bibinfo
  {volume} {2010}},\  (\bibinfo {year} {2010}{\natexlab{b}})},\ \Eprint
  {http://arxiv.org/abs/1006.4518} {arXiv:1006.4518} \BibitemShut {NoStop}%
\bibitem [{Note5()}]{Note5}%
  \BibitemOpen
  \bibinfo {note} {Note that $\protect \mathcal {E}$ is \protect \emph {not}
  the usual CP map $\protect \mathcal {F}$ arising from the isometry $\protect
  \mathcal {V}$, rather, it is the \protect \emph {inverse} or \protect \emph
  {transpose channel} of the CP map $\protect \mathcal {F}$.}\BibitemShut
  {Stop}%
\bibitem [{\citenamefont {Kadanoff}(1976)}]{kadanoff:1976a}%
  \BibitemOpen
  \bibfield  {author} {\bibinfo {author} {\bibfnamefont {L.~P.}\ \bibnamefont
  {Kadanoff}},\ }\href {\doibase 10.1016/0003-4916(76)90066-X} {\bibfield
  {journal} {\bibinfo  {journal} {Ann. Phys.}\ }\textbf {\bibinfo {volume}
  {100}},\ \bibinfo {pages} {359} (\bibinfo {year} {1976})}\BibitemShut
  {NoStop}%
\bibitem [{\citenamefont {Migdal}(1975{\natexlab{a}})}]{migdal:1975b}%
  \BibitemOpen
  \bibfield  {author} {\bibinfo {author} {\bibfnamefont {A.~A.}\ \bibnamefont
  {Migdal}},\ }\href@noop {} {\bibfield  {journal} {\bibinfo  {journal} {Zh.
  Eksp. Teor. Fiz.}\ }\textbf {\bibinfo {volume} {69}},\ \bibinfo {pages}
  {1457} (\bibinfo {year} {1975}{\natexlab{a}})}\BibitemShut {NoStop}%
\bibitem [{\citenamefont {Migdal}(1975{\natexlab{b}})}]{migdal:1975a}%
  \BibitemOpen
  \bibfield  {author} {\bibinfo {author} {\bibfnamefont {A.~A.}\ \bibnamefont
  {Migdal}},\ }\href@noop {} {\bibfield  {journal} {\bibinfo  {journal} {Zh.
  Eksp. Teor. Fiz.}\ }\textbf {\bibinfo {volume} {69}},\ \bibinfo {pages} {810}
  (\bibinfo {year} {1975}{\natexlab{b}})}\BibitemShut {NoStop}%
\bibitem [{Note6()}]{Note6}%
  \BibitemOpen
  \bibinfo {note} {This is a consequence of the coarse-grained state $\rho $
  not generally being invariant under ``non-diagonal'' gauge transformations
  $\rho \not =U_{\protect \{x\protect \}}\rho U_{\protect \{x'\protect
  \}}^\dagger $, where $U_{\protect \{x\protect \}}$ is the unitary that
  implements the gauge transformation parameterised by a set of gauge group
  elements $\protect \{x\protect \}$.}\BibitemShut {Stop}%
\bibitem [{\citenamefont {Livine}(2014)}]{livine_2014}%
  \BibitemOpen
  \bibfield  {author} {\bibinfo {author} {\bibfnamefont {E.~R.}\ \bibnamefont
  {Livine}},\ }\href {\doibase 10.1088/0264-9381/31/7/075004} {\bibfield
  {journal} {\bibinfo  {journal} {Class. Quantum Grav.}\ }\textbf {\bibinfo
  {volume} {31}},\ \bibinfo {pages} {075004} (\bibinfo {year} {2014})},\
  \Eprint {http://arxiv.org/abs/1310.3362} {arXiv:1310.3362} \BibitemShut
  {NoStop}%
\bibitem [{\citenamefont {Wesseling}(2004)}]{wesseling:2004a}%
  \BibitemOpen
  \bibfield  {author} {\bibinfo {author} {\bibfnamefont {P.}~\bibnamefont
  {Wesseling}},\ }\href@noop {} {\emph {\bibinfo {title} {An Introduction to
  {M}ultigrid {M}ethods}}}\ (\bibinfo  {publisher} {{R. T.} Edwards},\ \bibinfo
  {year} {2004})\BibitemShut {NoStop}%
\bibitem [{\citenamefont {Hastings}\ and\ \citenamefont
  {Wen}(2005)}]{hastings:2005a}%
  \BibitemOpen
  \bibfield  {author} {\bibinfo {author} {\bibfnamefont {M.~B.}\ \bibnamefont
  {Hastings}}\ and\ \bibinfo {author} {\bibfnamefont {X.-G.}\ \bibnamefont
  {Wen}},\ }\href {\doibase 10.1103/PhysRevB.72.045141} {\bibfield  {journal}
  {\bibinfo  {journal} {Phys. Rev. B}\ }\textbf {\bibinfo {volume} {72}},\
  \bibinfo {pages} {045141} (\bibinfo {year} {2005})},\ \Eprint
  {http://arxiv.org/abs/cond-mat/0503554} {arXiv:cond-mat/0503554} \BibitemShut
  {NoStop}%
\bibitem [{\citenamefont {Osborne}(2007)}]{osborne:2006a}%
  \BibitemOpen
  \bibfield  {author} {\bibinfo {author} {\bibfnamefont {T.~J.}\ \bibnamefont
  {Osborne}},\ }\href {\doibase 10.1103/PhysRevA.75.032321} {\bibfield
  {journal} {\bibinfo  {journal} {Phys. Rev. A}\ }\textbf {\bibinfo {volume}
  {75}},\ \bibinfo {pages} {032321} (\bibinfo {year} {2007})},\ \Eprint
  {http://arxiv.org/abs/quant-ph/0601019} {arXiv:quant-ph/0601019} \BibitemShut
  {NoStop}%
\bibitem [{\citenamefont {Bravyi}\ \emph {et~al.}(2010)\citenamefont {Bravyi},
  \citenamefont {Hastings},\ and\ \citenamefont {Michalakis}}]{bravyi:2010b}%
  \BibitemOpen
  \bibfield  {author} {\bibinfo {author} {\bibfnamefont {S.}~\bibnamefont
  {Bravyi}}, \bibinfo {author} {\bibfnamefont {M.~B.}\ \bibnamefont
  {Hastings}}, \ and\ \bibinfo {author} {\bibfnamefont {S.}~\bibnamefont
  {Michalakis}},\ }\href {\doibase 10.1063/1.3490195} {\bibfield  {journal}
  {\bibinfo  {journal} {J. Math. Phys.}\ }\textbf {\bibinfo {volume} {51}},\
  \bibinfo {pages} {093512} (\bibinfo {year} {2010})},\ \Eprint
  {http://arxiv.org/abs/1001.0344} {arxiv:1001.0344} \BibitemShut {NoStop}%
\bibitem [{\citenamefont {Bravyi}\ and\ \citenamefont
  {Hastings}(2011)}]{bravyi:2010c}%
  \BibitemOpen
  \bibfield  {author} {\bibinfo {author} {\bibfnamefont {S.}~\bibnamefont
  {Bravyi}}\ and\ \bibinfo {author} {\bibfnamefont {M.~B.}\ \bibnamefont
  {Hastings}},\ }\href {\doibase 10.1007/s00220-011-1346-2} {\bibfield
  {journal} {\bibinfo  {journal} {Comm. Math. Phys.}\ }\textbf {\bibinfo
  {volume} {307}},\ \bibinfo {pages} {609} (\bibinfo {year} {2011})},\ \Eprint
  {http://arxiv.org/abs/1001.4363} {arXiv:1001.4363} \BibitemShut {NoStop}%
\bibitem [{\citenamefont {Michalakis}\ and\ \citenamefont
  {Zwolak}(2013)}]{michalakis:2013a}%
  \BibitemOpen
  \bibfield  {author} {\bibinfo {author} {\bibfnamefont {S.}~\bibnamefont
  {Michalakis}}\ and\ \bibinfo {author} {\bibfnamefont {J.~P.}\ \bibnamefont
  {Zwolak}},\ }\href {\doibase 10.1007/s00220-013-1762-6} {\bibfield  {journal}
  {\bibinfo  {journal} {Comm. Math. Phys.}\ }\textbf {\bibinfo {volume}
  {322}},\  (\bibinfo {year} {2013})},\ \Eprint
  {http://arxiv.org/abs/1109.1588} {arXiv:1109.1588} \BibitemShut {NoStop}%
\bibitem [{\citenamefont {Lieb}\ and\ \citenamefont
  {Robinson}(1972)}]{lieb:1972a}%
  \BibitemOpen
  \bibfield  {author} {\bibinfo {author} {\bibfnamefont {E.~H.}\ \bibnamefont
  {Lieb}}\ and\ \bibinfo {author} {\bibfnamefont {D.~W.}\ \bibnamefont
  {Robinson}},\ }\href {\doibase 10.1007/BF01645779} {\bibfield  {journal}
  {\bibinfo  {journal} {Comm. Math. Phys.}\ }\textbf {\bibinfo {volume} {28}},\
  \bibinfo {pages} {251} (\bibinfo {year} {1972})}\BibitemShut {NoStop}%
\bibitem [{\citenamefont {Hastings}\ and\ \citenamefont
  {Koma}(2006)}]{hastings:2005b}%
  \BibitemOpen
  \bibfield  {author} {\bibinfo {author} {\bibfnamefont {M.~B.}\ \bibnamefont
  {Hastings}}\ and\ \bibinfo {author} {\bibfnamefont {T.}~\bibnamefont
  {Koma}},\ }\href {\doibase 10.1007/s00220-006-0030-4} {\bibfield  {journal}
  {\bibinfo  {journal} {Comm. Math. Phys.}\ }\textbf {\bibinfo {volume}
  {265}},\ \bibinfo {pages} {781} (\bibinfo {year} {2006})},\ \Eprint
  {http://arxiv.org/abs/math-ph/0507008} {arXiv:math-ph/0507008} \BibitemShut
  {NoStop}%
\bibitem [{\citenamefont {Nachtergaele}\ and\ \citenamefont
  {Sims}(2010)}]{nachtergaele:2010b}%
  \BibitemOpen
  \bibfield  {author} {\bibinfo {author} {\bibfnamefont {B.}~\bibnamefont
  {Nachtergaele}}\ and\ \bibinfo {author} {\bibfnamefont {R.}~\bibnamefont
  {Sims}},\ }in\ \href@noop {} {\emph {\bibinfo {booktitle} {Entropy and the
  quantum}}},\ \bibinfo {series} {Contemp. Math.}, Vol.\ \bibinfo {volume}
  {529}\ (\bibinfo  {publisher} {Amer. Math. Soc.},\ \bibinfo {address}
  {Providence, RI},\ \bibinfo {year} {2010})\ pp.\ \bibinfo {pages}
  {141--176}\BibitemShut {NoStop}%
\bibitem [{\citenamefont {Pr\'emont-Schwarz}\ \emph {et~al.}(2010)\citenamefont
  {Pr\'emont-Schwarz}, \citenamefont {Hamma}, \citenamefont {Klich},\ and\
  \citenamefont {Markopoulou-Kalamara}}]{premont-schwarz:2010a}%
  \BibitemOpen
  \bibfield  {author} {\bibinfo {author} {\bibfnamefont {I.}~\bibnamefont
  {Pr\'emont-Schwarz}}, \bibinfo {author} {\bibfnamefont {A.}~\bibnamefont
  {Hamma}}, \bibinfo {author} {\bibfnamefont {I.}~\bibnamefont {Klich}}, \ and\
  \bibinfo {author} {\bibfnamefont {F.}~\bibnamefont {Markopoulou-Kalamara}},\
  }\href {\doibase 10.1103/PhysRevA.81.040102} {\bibfield  {journal} {\bibinfo
  {journal} {Phys. Rev. A}\ }\textbf {\bibinfo {volume} {81}},\ \bibinfo
  {pages} {040102} (\bibinfo {year} {2010})},\ \Eprint
  {http://arxiv.org/abs/0912.4544} {arXiv:0912.4544} \BibitemShut {NoStop}%
\bibitem [{\citenamefont {Osborne}(2006)}]{osborne:2005d}%
  \BibitemOpen
  \bibfield  {author} {\bibinfo {author} {\bibfnamefont {T.~J.}\ \bibnamefont
  {Osborne}},\ }\href {\doibase 10.1103/PhysRevLett.97.157202} {\bibfield
  {journal} {\bibinfo  {journal} {Phys. Rev. Lett.}\ }\textbf {\bibinfo
  {volume} {97}},\ \bibinfo {pages} {157202} (\bibinfo {year} {2006})},\
  \Eprint {http://arxiv.org/abs/quant-ph/0508031} {arXiv:quant-ph/0508031}
  \BibitemShut {NoStop}%
\bibitem [{\citenamefont {Peskin}\ and\ \citenamefont
  {Schroeder}(1995)}]{peskin:1995a}%
  \BibitemOpen
  \bibfield  {author} {\bibinfo {author} {\bibfnamefont {M.~E.}\ \bibnamefont
  {Peskin}}\ and\ \bibinfo {author} {\bibfnamefont {D.~V.}\ \bibnamefont
  {Schroeder}},\ }\href@noop {} {\emph {\bibinfo {title} {An introduction to
  quantum field theory}}}\ (\bibinfo  {publisher} {Westview Press},\ \bibinfo
  {year} {1995})\BibitemShut {NoStop}%
\bibitem [{\citenamefont {{Arnold}}\ \emph {et~al.}(2003)\citenamefont
  {{Arnold}}, \citenamefont {{Bunk}}, \citenamefont {{Lippert}},\ and\
  \citenamefont {{Schilling}}}]{arnold:2003a}%
  \BibitemOpen
  \bibfield  {author} {\bibinfo {author} {\bibfnamefont {G.}~\bibnamefont
  {{Arnold}}}, \bibinfo {author} {\bibfnamefont {B.}~\bibnamefont {{Bunk}}},
  \bibinfo {author} {\bibfnamefont {T.}~\bibnamefont {{Lippert}}}, \ and\
  \bibinfo {author} {\bibfnamefont {K.}~\bibnamefont {{Schilling}}},\ }\href
  {\doibase 10.1016/S0920-5632(03)01704-3} {\bibfield  {journal} {\bibinfo
  {journal} {Nucl. Phys. B Proc. Suppl.}\ }\textbf {\bibinfo {volume} {119}},\
  \bibinfo {pages} {864} (\bibinfo {year} {2003})},\ \Eprint
  {http://arxiv.org/abs/hep-lat/0210010} {arXiv:hep-lat/0210010} \BibitemShut
  {NoStop}%
\bibitem [{\citenamefont {Milsted}(2016)}]{milsted_matrix_2016}%
  \BibitemOpen
  \bibfield  {author} {\bibinfo {author} {\bibfnamefont {A.}~\bibnamefont
  {Milsted}},\ }\href {\doibase 10.1103/PhysRevD.93.085012} {\bibfield
  {journal} {\bibinfo  {journal} {Phys. Rev. D}\ }\textbf {\bibinfo {volume}
  {93}},\ \bibinfo {pages} {085012} (\bibinfo {year} {2016})},\ \Eprint
  {http://arxiv.org/abs/1507.06624} {arXiv:1507.06624} \BibitemShut {NoStop}%
\bibitem [{Note7()}]{Note7}%
  \BibitemOpen
  \bibinfo {note} {This terminology was suggested to us by Vaughan
  Jones.}\BibitemShut {Stop}%
\bibitem [{\citenamefont {Lang}(2002)}]{lang:2002a}%
  \BibitemOpen
  \bibfield  {author} {\bibinfo {author} {\bibfnamefont {S.}~\bibnamefont
  {Lang}},\ }\href@noop {} {\emph {\bibinfo {title} {Algebra}}},\ \bibinfo
  {edition} {3rd}\ ed.,\ \bibinfo {series} {Graduate Texts in Mathematics},
  Vol.\ \bibinfo {volume} {211}\ (\bibinfo  {publisher} {Springer-Verlag},\
  \bibinfo {address} {New York},\ \bibinfo {year} {2002})\ pp.\ \bibinfo
  {pages} {xvi+914}\BibitemShut {NoStop}%
\bibitem [{\citenamefont {Bimonte}\ \emph {et~al.}(1996)\citenamefont
  {Bimonte}, \citenamefont {Ercolessi}, \citenamefont {Landi}, \citenamefont
  {Lizzi}, \citenamefont {Sparano},\ and\ \citenamefont
  {Teotonio-Sobrinho}}]{bimonte_lattices_1996}%
  \BibitemOpen
  \bibfield  {author} {\bibinfo {author} {\bibfnamefont {G.}~\bibnamefont
  {Bimonte}}, \bibinfo {author} {\bibfnamefont {E.}~\bibnamefont {Ercolessi}},
  \bibinfo {author} {\bibfnamefont {G.}~\bibnamefont {Landi}}, \bibinfo
  {author} {\bibfnamefont {F.}~\bibnamefont {Lizzi}}, \bibinfo {author}
  {\bibfnamefont {G.}~\bibnamefont {Sparano}}, \ and\ \bibinfo {author}
  {\bibfnamefont {P.}~\bibnamefont {Teotonio-Sobrinho}},\ }\href {\doibase
  10.1016/S0393-0440(95)00063-1} {\bibfield  {journal} {\bibinfo  {journal} {J.
  Geom. Phys.}\ }\textbf {\bibinfo {volume} {20}},\ \bibinfo {pages} {318}
  (\bibinfo {year} {1996})},\ \Eprint {http://arxiv.org/abs/hep-th/9507147}
  {arXiv:hep-th/9507147} \BibitemShut {NoStop}%
\bibitem [{\citenamefont {Ashtekar}\ and\ \citenamefont
  {Isham}(1992)}]{ashtekar_representations_1992}%
  \BibitemOpen
  \bibfield  {author} {\bibinfo {author} {\bibfnamefont {A.}~\bibnamefont
  {Ashtekar}}\ and\ \bibinfo {author} {\bibfnamefont {C.~J.}\ \bibnamefont
  {Isham}},\ }\href {\doibase 10.1088/0264-9381/9/6/004} {\bibfield  {journal}
  {\bibinfo  {journal} {Classical and Quantum Gravity}\ }\textbf {\bibinfo
  {volume} {9}},\ \bibinfo {pages} {1433} (\bibinfo {year} {1992})},\ \Eprint
  {http://arxiv.org/abs/hep-th/9202053} {arXiv:hep-th/9202053} \BibitemShut
  {NoStop}%
\bibitem [{\citenamefont {Ashtekar}\ and\ \citenamefont
  {Lewandowski}(1993)}]{ashtekar_representation_1993}%
  \BibitemOpen
  \bibfield  {author} {\bibinfo {author} {\bibfnamefont {A.}~\bibnamefont
  {Ashtekar}}\ and\ \bibinfo {author} {\bibfnamefont {J.}~\bibnamefont
  {Lewandowski}},\ }\href@noop {} {\  (\bibinfo {year} {1993})},\ \Eprint
  {http://arxiv.org/abs/gr-qc/9311010} {arXiv:gr-qc/9311010} \BibitemShut
  {NoStop}%
\bibitem [{\citenamefont {Dittrich}(2012)}]{dittrich_discrete_2012}%
  \BibitemOpen
  \bibfield  {author} {\bibinfo {author} {\bibfnamefont {B.}~\bibnamefont
  {Dittrich}},\ }\href {\doibase 10.1088/1367-2630/14/12/123004} {\bibfield
  {journal} {\bibinfo  {journal} {New J. Phys.}\ }\textbf {\bibinfo {volume}
  {14}},\ \bibinfo {pages} {123004} (\bibinfo {year} {2012})},\ \Eprint
  {http://arxiv.org/abs/1205.6127} {arXiv:1205.6127} \BibitemShut {NoStop}%
\bibitem [{\citenamefont {Jones}(2014)}]{jones_unitary_2014}%
  \BibitemOpen
  \bibfield  {author} {\bibinfo {author} {\bibfnamefont {V.~F.~R.}\
  \bibnamefont {Jones}},\ }\href@noop {} {\  (\bibinfo {year} {2014})},\
  \Eprint {http://arxiv.org/abs/1412.7740} {arXiv:1412.7740} \BibitemShut
  {NoStop}%
\end{thebibliography}
\end{document}